\documentclass[twocolumn]{aastex63}

\submitjournal{AJ}

\shorttitle{Looking at Blazar Light Curve Periodicities}
\shortauthors{Covino et al.}
\graphicspath{{./}{figures/}}

\begin{document}

\title{Looking at Blazar Light Curve Periodicities with Gaussian Processes}

\correspondingauthor{Stefano Covino}
\email{stefano.covino@inaf.it}

\author[0000-0001-9078-5507]{Stefano Covino}
\affiliation{INAF / Brera Astronomical Observatory, via Bianchi 46, 23807, Merate (LC), Italy}

\author{Marco Landoni}
\affiliation{INAF / Brera Astronomical Observatory, via Bianchi 46, 23807, Merate (LC), Italy}

\author{Angela Sandrinelli}
\affiliation{INAF / Brera Astronomical Observatory, via Bianchi 46, 23807, Merate (LC), Italy}

\author{Aldo Treves}
\affiliation{INAF / Brera Astronomical Observatory, via Bianchi 46, 23807, Merate (LC), Italy}
\affiliation{Universit\`a degli Studi dell'Insubria, Via Valleggio 11, 22100 Como, Italy}



\begin{abstract}

Temporal analysis of blazar flux is a powerful tool to draw inferences about the emission processes and physics of these sources. In the most general case, the available light curves are irregularly sampled and influenced by gaps, and in addition are also affected by correlated noise, making their analysis complicated. Gaussian processes may offer a viable tool to assess the statistical significance of proposed periods in light curves characterized by any sampling and noise pattern.
We infer the significance of the periods proposed in the literature for two well known blazars with multiple claims of possible year-long periodicity: PG\,1553+113 and PKS\,2155-304, in the high-energy and optical bands.  Adding a periodic component to the modeling gives a better statistical description of the analyzed light curves. The improvement is rather solid for PG\,1553+113, both at high energies and in the optical, while for PKS\,2155-304 at high energies the improvement is not yet strong enough to allow cogent claims, and no evidence for periodicity emerged by the analysis in the optical.
Modeling a light curve by means of Gaussian processes, in spite of being relatively computationally demanding, allows us to derive a wealth of information about the data under study and suggests an original analysis framework for light curves of astrophysical interest.

\end{abstract}



\section{Introduction}
\label{sec:intro}

Blazars \citep[e.g.,][]{Urry2012} are one of the most frequent targets for monitoring programs at essentially any wavelength because of their variability, from the radio band to the highest energies. A natural outcome of this intense monitoring is the availability of long time series that offer a treasure of possible information about this class of sources \citep[e.g.,][]{Ryanetal2019}. One of the most interesting topics that can be addressed with well sampled time series is the identification of possible periodic or quasi-periodic behaviors. Several authors have indeed proposed quasi-periodic oscillations (QPO) for blazars at various levels of statistical significance \citep[e.g.,][]{Lehto&Valtonen1996,Zhangetal2014,Sandrinellietal2014,Ackermannetal2015,Cutinietal2016,Sandrinellietal2016a,Sandrinellietal2016b,Stamerraetal2016,Sandrinellietal2017,Covinoetal2017,Cavaliereetal2017,ProkhorovMoraghan2017,Zhangetal2017a,Zhangetal2017b,Zhangetal2017c,Tavanietal2018,Sandrinellietal2018,Bhatta2019,Covinoetal2019,Rieger2019,Chevalieretal2019,AitBenkhalietal2019,Bhatta&Dhital2019}.

This interest is not misplaced, since QPOs in blazars could be powerful diagnostics of different phenomena associated to blazar phenomenologies \citep[e.g.,][]{Lindforsetal2016}. Interpretation of possible QPOs in blazars has recently attracted the attention of several works \citep[e.g.,][]{Cavaliereetal2017,Sobacchietal2017,Capronietal2017,Holgadoetal2018,Cavaliereetal2019,Licoetal2020}, mainly motivated by the various claims for periodicities proposed in the last few years. One picture of interest envisages the QPOs as due to a supermassive binary black-hole systems. This might directly introduce periodicities in the observed emissions \citep[e.g.,][]{Lehto&Valtonen1996,Sandrinellietal2014,Grahametal2015} or indirectly through precession of the whole system \citep{Doganetal2015}. On the other hand, QPOs, possibly of transient nature, could also be generated by instabilities in the relativistic jets or in the accretion disks \citep{Camenzind&Krockenberger1992,Marscher2014,Raiterietal2017}.

\section{Time series analysis}
\label{sec:tsa}

The analysis of time series is a fundamental tool in modern astrophysics \citep[e.g.,][]{Vaughan2013}. Given the large variety of astronomical data we can deal with, it is natural that many different techniques can be applied. Without any claim of completeness, two general scenarios can be depicted: those to treat evenly spaced data and those to cope with the irregular sampling as is the case with
astronomical observations. Evenly sampled data offer the remarkable advantage to allow analyses based on a well developed set of procedures and theorems to interpret their results \citep[e.g.,][for a comprehensive review]{vanderKlis1989}. The case of irregularly sampled data can be dealt with the recipes developed by \citet{Lomb1976} and \citet{Scargle1982}. The Lomb-Scargle (LS) periodogram offers a rigorous solution to the problem of detecting periodic signals in noisy time series \citep{Bretthorst2003,VanderPlas2018}. However, it is also known to produce distorted versions of the true periodogram and alternative approaches have also been proposed \citep[e.g.,][]{Vio&Andreani2018}. Moreover, decomposing a light curve into harmonic series unavoidably implies a particular sensitivity to quasi-sinusoidal variations. In order to cope with periodicities with any functional form, non-parametric analysis tools have been developed \citep[e.g.,][]{Stellingwerf1978,Schwarzenberg-Czerny1997,Huijseetal2018}. 

Independently of the particular analysis recipe, the problem of assessing the significance of any possible periodicity in a light curve requires one to evaluate the probability that at the frequency of interest the measured power (or any other adopted periodicity indicator) is not due, to a given confidence level, to random fluctuations. This assessment depends critically on the noise affecting the data. In case we can assume the noise is ``white", i.e. independent of the frequency, it is possible to compute exactly the significance level of any observed power. The most general case, red or correlated noise \citep{Press1978,Milotti2002,Milotti2007}, is considerable more difficult and still an open problem. For evenly spaced data several approaches have been suggested \citep[e.g.,][]{Vaughan2010,Barret&Vaughn2012,Guidorzietal2016,Vaughanetal2016}, while for the more common unequally spaced observations only indirect procedures can be applied. There are anyway several problems to deal with. Again, with no claim of completeness for this widely debated topic, we can mention the unknown distribution of periodogram peaks, the poor measurement of population variance \citep{Koen1990} and the independence of the powers for a periodogram at the various frequencies. This is guaranteed only when data are evenly spaced and the periodogram is computed at the Fourier frequencies, or on any other ortho-normal frequency grid. Under this condition, a formal fit of the noise functional form is possible and inferences can be derived once the noise has been properly modeled. Otherwise, a procedure often followed consists in generating a large number of dense, long, and highly sampled simulated light-curves given a set of possible noise models \citep[or noise model parameters,][]{Uttleyetal2002}. Then, light-curves with the same sampling of the observed curve are obtained and statistics about the power of the derived periodograms are obtained \citep[e.g.,][]{Ackermannetal2015,Bhatta2019,Grahametal2015,Tavanietal2018,Nilssonetal2018}. Other approaches are of course possible, depending on the specific data set in analysis. As a matter of fact, the fundamental problem of assessing the significance of peaks in astronomical periodograms have understandably always received great attention in literature  \citep[e.g.][]{Baluev2008,Suvegesetal2015,Haraetal2017,Sulisetal2017}.

An alternative procedure is to derive information about possible periodic behaviors working in the time rather than frequency domain \citep[e.g.,][]{Li&Wang2018,Feigelsonetal2019}. While often computationally expensive, working in the time domain presents several advantages. First of all, one can apply inferences with data with uncertainties normally distributed. This is not (in general) the case in the frequency domain \citep{Israel&Stella1996}. In addition phenomena affecting analysis in the frequency domain as leakage, aliasing, etc. are unimportant or much easier to deal with. The inferences are also often made less ambiguous since one does not have to rely on asymptotic behaviors, statistical tests strictly holding only for infinite time-series, or hypotheses of stationarity \citep{Kwiatkowskietal1992}.

An effective approach to time series modeling in the time domain has been developed to check if the observed data can be generated by autoregressive processes \citep{Koen2005,Kellyetal2009,Kellyetal2014,Lenoir&Crucifix2018,Takataetal2018,Kovacevicetal2019,Feigelsonetal2019,Elorrietaetal2019}. It is indeed well known that if the noise is correlated then spurious and transients quasi-periodicities can be observed \citep[e.g.,][]{Vaughanetal2016}. A partly different procedure, that is  becoming increasing popular in the astronomical literature \citep[e.g.,][]{Brewer&Stello2009,Wangetal2012,Haywoodetal2014,Ivezicetal2014,Vanderburgetal2015,Rajpauletal2015,Lugeretal2016,Karamanavis2017,Foreman-Mackeyetal2017,Kovacevicetal2019,Wilkins2019,Pereiraetal2019,Jesusetal2019,Chuaetal2019} and more generally in Bayesian signal estimation, is time series analysis based on Gaussian Processes \citep[GP, ][]{Rasmussen&Williams2006,Robertsetal2012,Durrandeetal2016,Tobaretal2015,Littlefairetal2017,Tobar2018,Angusetal2018}. 
GP analysis is intrinsically a Bayesian technique, i.e. prior information that encapsulates our assumptions on the analyzed time series (such as smoothness, stationarity or periodicity) is adopted. Then, this is updated with the information provided by the observed data via a given likelihood function. And finally, a posterior distribution of the derived parameters can be used for any inference. GP are a generalization of multivariate Gaussian distributions of variables and offer a very flexible framework for modelling unknown functions by non-parametric models.  A key component of the analysis is the kernel or covariance function. Given any arbitrary pair of observations, the kernel defines the degree of similarity between the observed values. There can be a plethora of possible kernel functions (squared-exponential, Mat\'ern, etc.), although in practice, in most cases, just a few basic functions are used \citep{Rasmussen&Williams2006}. Kernel functions drive the degree of smoothness of the observed light-curves, and can also identify periodic behaviors and define an important connection between autoregressive time series methods and GP analysis \citep{Rasmussen&Williams2006,Durrandeetal2016,Foreman-Mackeyetal2017}.

\section{Methods}
\label{sec:meth}

Our goal is to quantify the significance of possible periodicites in blazar light-curves applying a procedure that allows us to draw inferences essentially independently of the sampling scheme of the analyzed (possibly multi-dimensional) data. 

The choice of a specific covariance function can be often relatively arbitrary. A frequent choice, as in \citet{Angusetal2018}, is the Square Exponential (SE) covariance function \citep{Rasmussen&Williams2006}:
\begin{equation}
k_{r} = A \exp \left(-\frac{r^2}{2L^2} \right),
\label{eq:SE}
\end{equation}
where $A>0$ is the amplitude, $L$ is the length scale of the exponential decay, and $r = (t_i-t_j)$ is the time separation between data points. This is a stationary kernel since it depends only on the data separation. The choice of the SE kernel is mainly driven by its simplicity, depending only on two parameters, $A$ and $L$, although there are alternatives with the same number of parameters. In general, if $L$ is large, the correlation between two data points largely separated will be stronger. 

The SE kernel could be seen as a special case of the more general Mat\'ern covariance function family \citep{Rasmussen&Williams2006,Durrandeetal2016,Foreman-Mackeyetal2017}. The Mat\'ern kernel functions are characterized by a parameter, $\nu$, that drives the degree of ``smoothness" of the kernel. Functions with $\nu = p - 1/2$ are the discrete time equivalent to AR (autoregressive) processes of order $p$. For $\nu \to +\infty$ the kernel becomes the SE covariance function while, with $\nu = \frac{1}{2}$, it simplifies to the Absolute Exponential (AE) kernel (i.e. the covariance of a Ornstein-Uhlenbeck process):
\begin{equation}
k_{r} = A \exp \left(-\frac{r}{L} \right).
\label{eq:AE}
\end{equation}

Finally, another interesting and frequently used stationary kernel is the Rational Quadratic (RQ) covariance function \citep{Rasmussen&Williams2006}:
\begin{equation}
k_{r} = A \left[1 + \left(\frac{r^2}{2\alpha L^2} \right)^{-\alpha}\right],
\label{eq:RQ}
\end{equation}
with $\alpha$ strictly positive. This kernel can be seen as a scale mixture (i.e. an infinite sum) of SE covariance functions with different characteristic length-scales drawn from a {\tt gamma} distribution. The limit of the RQ covariance for $\alpha \to +\infty$ is indeed just the SE covariance function with characteristic length-scale $L$.

In general, any covariance function mentioned above allows a satisfactory modeling of the datasets. However, having to deal with time series analysis requires some more care. The light curves considered in this work are characterized by a power spectral density (PSD) modeled by a power-law (PL) or a broken PL \citep[see Sect.\,\ref{sec:data} and, e.g., ][]{Nilssonetal2018,Covinoetal2019} and therefore it is important that the adopted covariance functions provide an adequate description of the data and their PSDs. As reported by \citet{Wilkins2019}, the SE kernel does not correctly describe the PSD of the data, in particular at low frequencies. On the contrary, the AE and RQ kernels provide better results. The best choice depends on the specific PSD functional form. The multiple length-scale sensitivity of the RQ kernel makes it an interesting choice for blazar light curves, i.e. characterized by a complex variability pattern, however the SE and AE kernels need a lower number of free parameters. We considered in our analyses both the AE and the RQ kernels, as discussed in Sect.\,\ref{sec:stat}.

Different families of kernels are needed to describe periodic behaviors. The simplest possibility is ``the cosine" (CS) kernel:
\begin{equation}
k_{r} = A \cos (2 \pi r / P),
\label{eq:cos}
\end{equation}
where again $A>0$ is the amplitude and $r$ is the separation between data points. $P$ is the period in the data.
A more flexible and also widely used covariance function is the ``exp-sine-squared" (ESS) kernel \citep{Rasmussen&Williams2006}:
\begin{equation}
k_{r} = A \exp \left[-\Gamma^2 \sin^2\left(\frac{\pi r}{P}\right) \right],
\label{eq:ESS}
\end{equation}
where, if the additional parameter $\Gamma$ is large, points separated by a period are strongly correlated, while the correlation is looser if $\Gamma$ becomes small.

The ESS kernel offers a larger flexibility in modeling quasi-periodic phenomena, yet the CS kernel also allows negative covariances, a feature often present in case of periodic behaviors and characterizing the auto-correlation functions of the data considered in this study (see Section\,\ref{sec:qp}). 

The product or the sum of two (or more) kernel functions is still a legitimate kernel function \citep{Rasmussen&Williams2006}. The sum and product of covariance functions reflect two different scenarios. A sum of two kernels gives higher values when the first ``or" the second operands support high correlation, while for the product this occurs when the first ``and" the second operands are both giving high correlation. In the following we report results both for the sum and the product and adopt a kernel derived by the combination of the AE or RQ and the CS covariance functions. 

The problem of determining the possible presence of a periodic behavior in a light-curve can thus be converted to a plain model comparison in a Bayesian framework \citep[e.g., ][]{Kass&Raftery1995,Trotta2008,Ivezicetal2014,Andreon&Weaver2015}, i.e. fitting the data with a stationary kernel and a more complex model with a periodic component. We also did not define any prior on the mean function, in order not to bias our results assuming a given functional form. The procedure we followed consists in initially maximizing the likelihood function by a non-linear optimization algorithm \citep[e.g. the Nelder-Mead or the L-BFGS-B algorithms, ][]{Gao&Han2012,Byrdetal1995} and integrating the posterior probability density of the parameters of our models by a Markov Chain Monte Carlo \citep[MCMC,][]{Hogg&Foreman2018} based on the ``affine-invariant Hamiltonian" and the ``parallel-tempering ensemble" algorithms \citep{Foreman-Mackeyetal2013}. We started the chains from small Gaussian balls centered on the best fit values. The first third of each chain (the ``burn-in phase") was discarded and we checked that a stationary distribution was reached \citep{Sharma2017}. Model comparison could be carried out by evaluating the Bayesian Information Criterion \citep[BIC,][]{Schwarz1978}, which is simple to compute but requires that the posterior distribution of the parameters is essentially Gaussian, often an assumption not satisfied. We therefore  carried out model comparison by a full computation of the Bayes factors \citep{Ivezicetal2014}, typically much more demanding computationally, but not requiring any special assumption \citep[see also][]{Liddle2007}. We computed the Bayes factors, following \citet{Littlefairetal2017}, by the so-called ``thermodynamic integration" \citep{Goggans&Chi2004}, which indeed offers an effective compromise between accuracy and computational complexity.

The model comparison to assess whether the introduction of a periodic term is preferred compared to stationary covariance functions is then carried out leaving the GP regression free to identify possible periods (within a given large range) that can then be evaluated analysing the posterior distribution of periods obtained after the analysis (Sect.\,\ref{sec:res}). 

Finally, the significance of the introduction of a periodic component in the modeling of the light curve has typically to be corrected for a trial factor, i.e. the number of independent frequencies that are ``tried" (explicitly or implicitly) during the analysis. There is a large literature about this topic. The number of independent frequencies depends on the length of the time series and on the sampling. This is not typically a problem with a simple solution, but approximate estimates are often adequate for most practical purposes \citep[e.g.][]{Horne&Baliunas1986,Schwarzenberg-Czerny1998,Cumming2004,Baluev2008,Frescuraetal2008,Zechmeister&Kurster2009,Baluev2013,Suveges2014}.

However, the computation of the Bayes factor already includes the multi-trial correction, since it is embedded in the integration on the allowed parameter space defined by the priors on the analysis \citep[e.g.,][]{Gelman&Tuerlinckx2000,Trotta2007,Gelmanetal2012}. Effectively, the Bayes factor model selection takes into proper account all the information provided by the data.

Software tools and packages used throughout the present analyses are listed in Appendix\,\ref{ap:soft}.

\section{Data}
\label{sec:data}

\begin{figure*}
\begin{tabular}{cc}
\includegraphics[width=8cm]{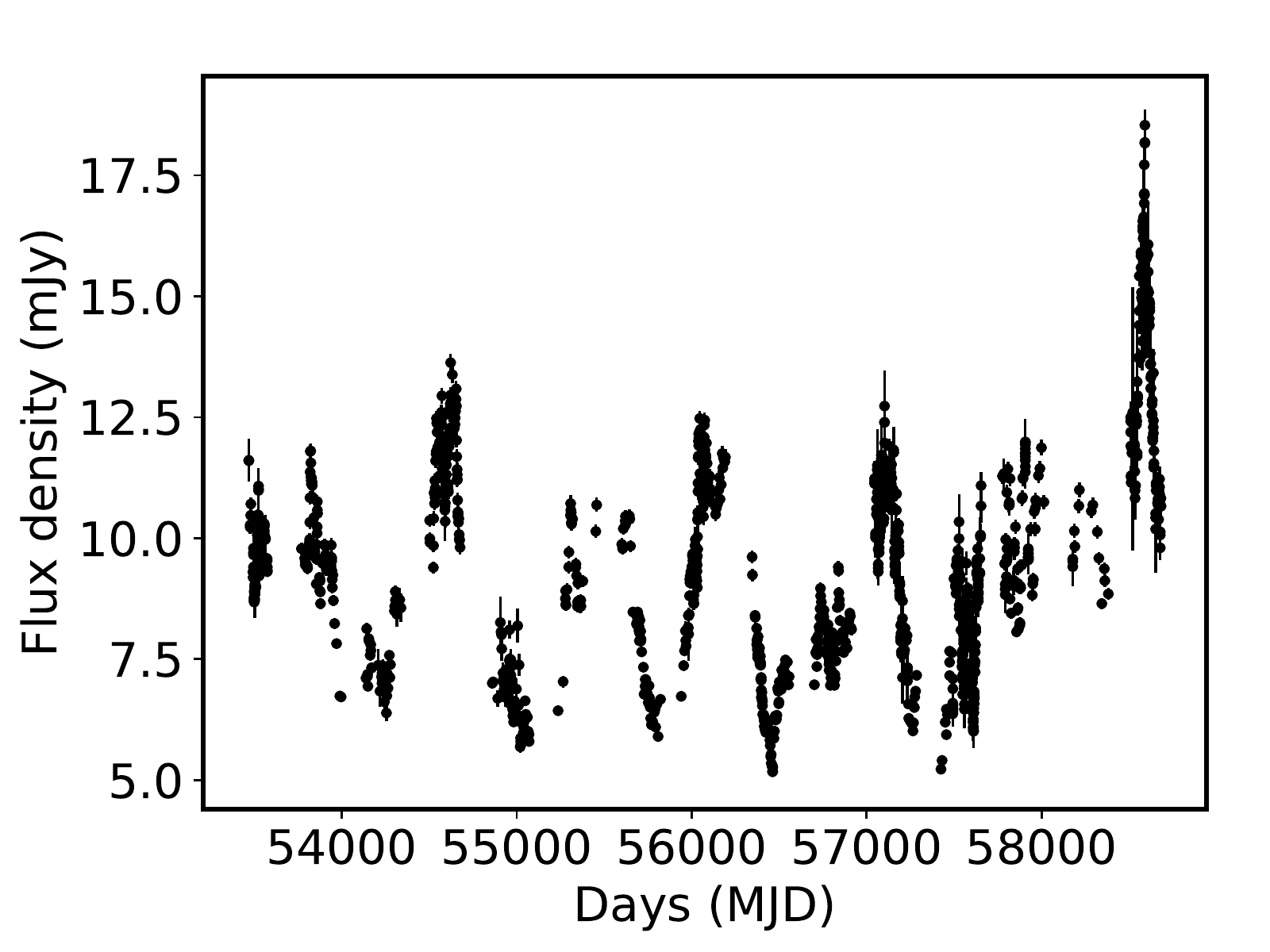} & \includegraphics[width=8cm]{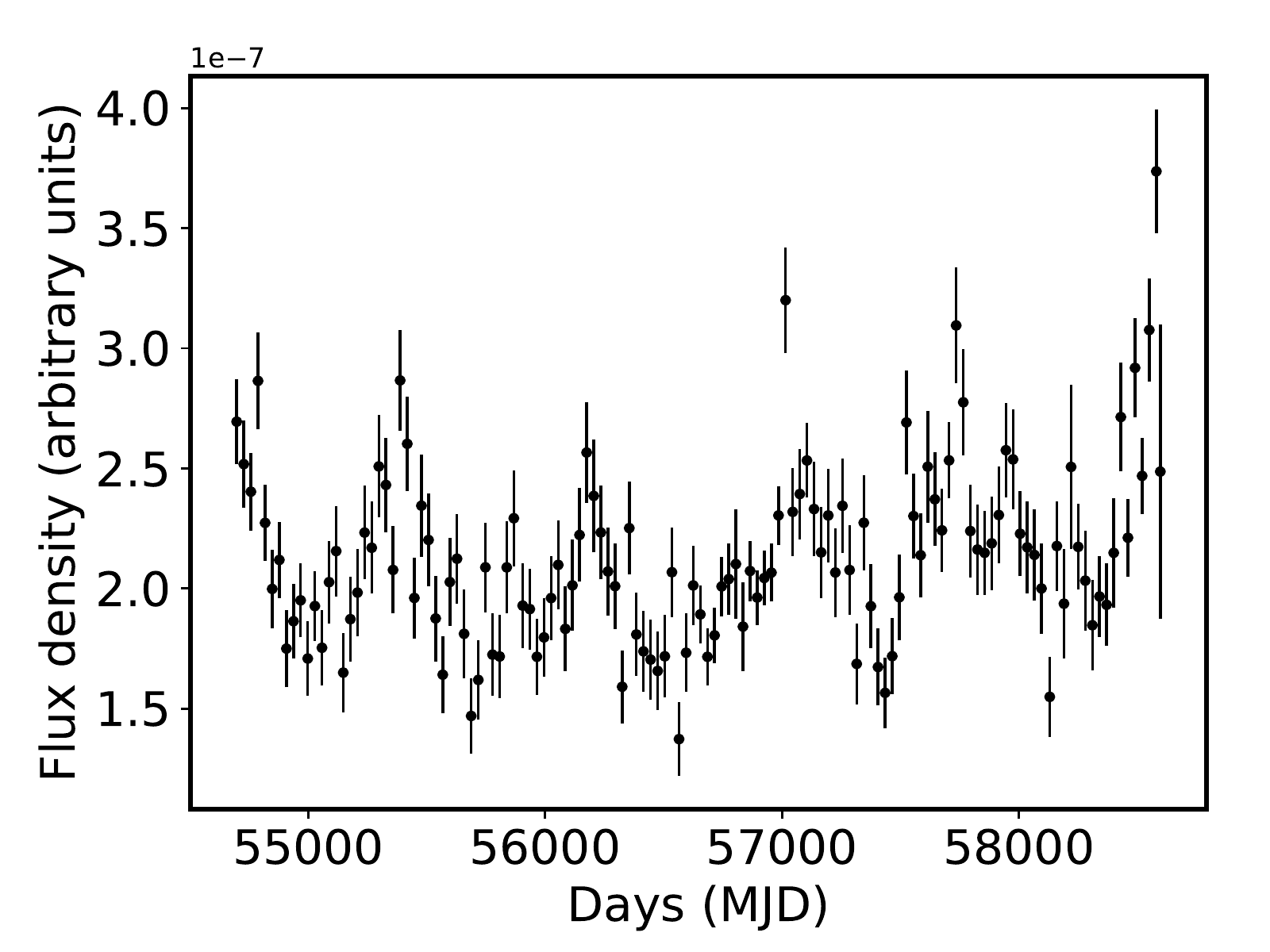} \\
\includegraphics[width=8cm]{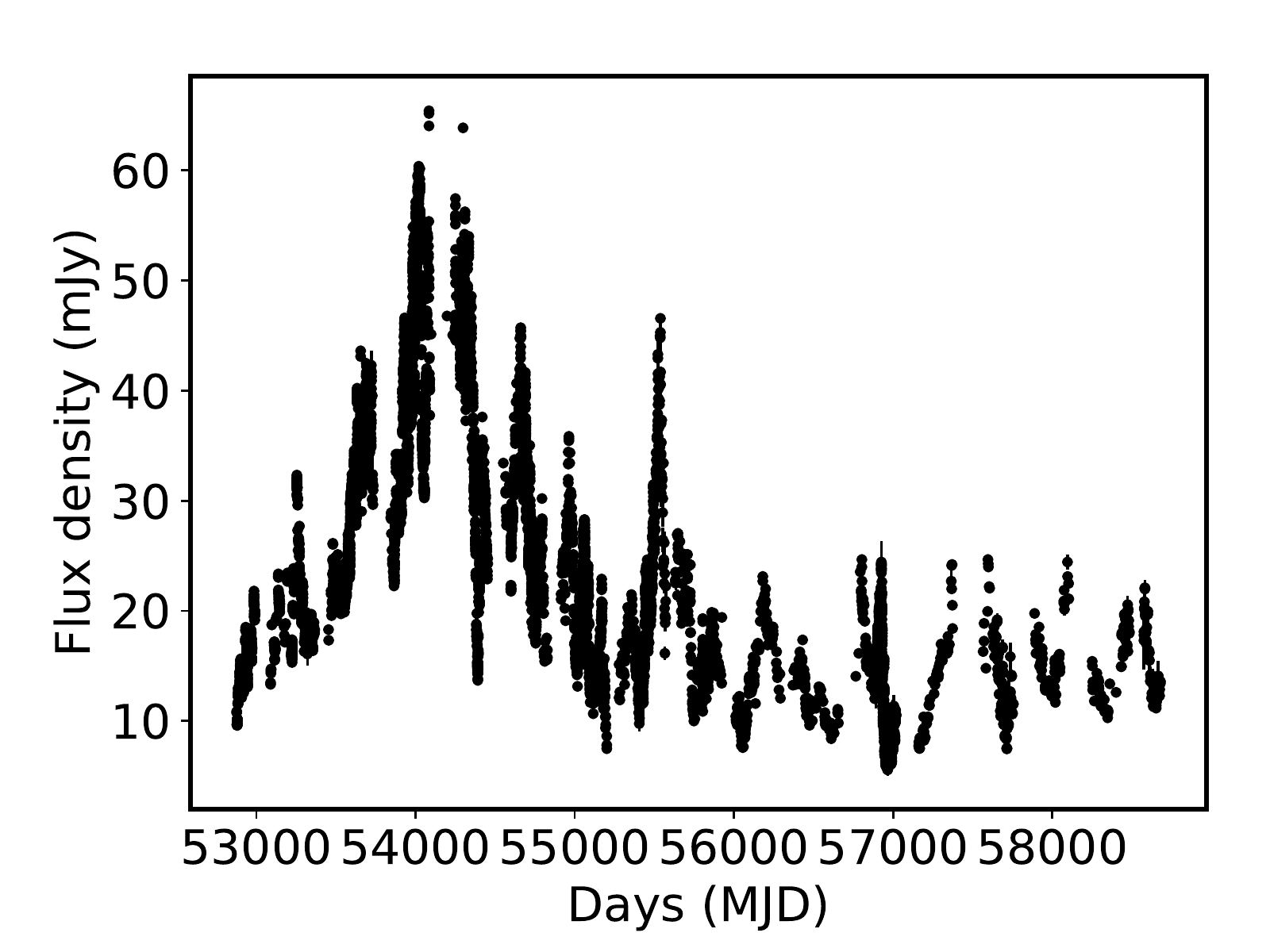} & \includegraphics[width=8cm]{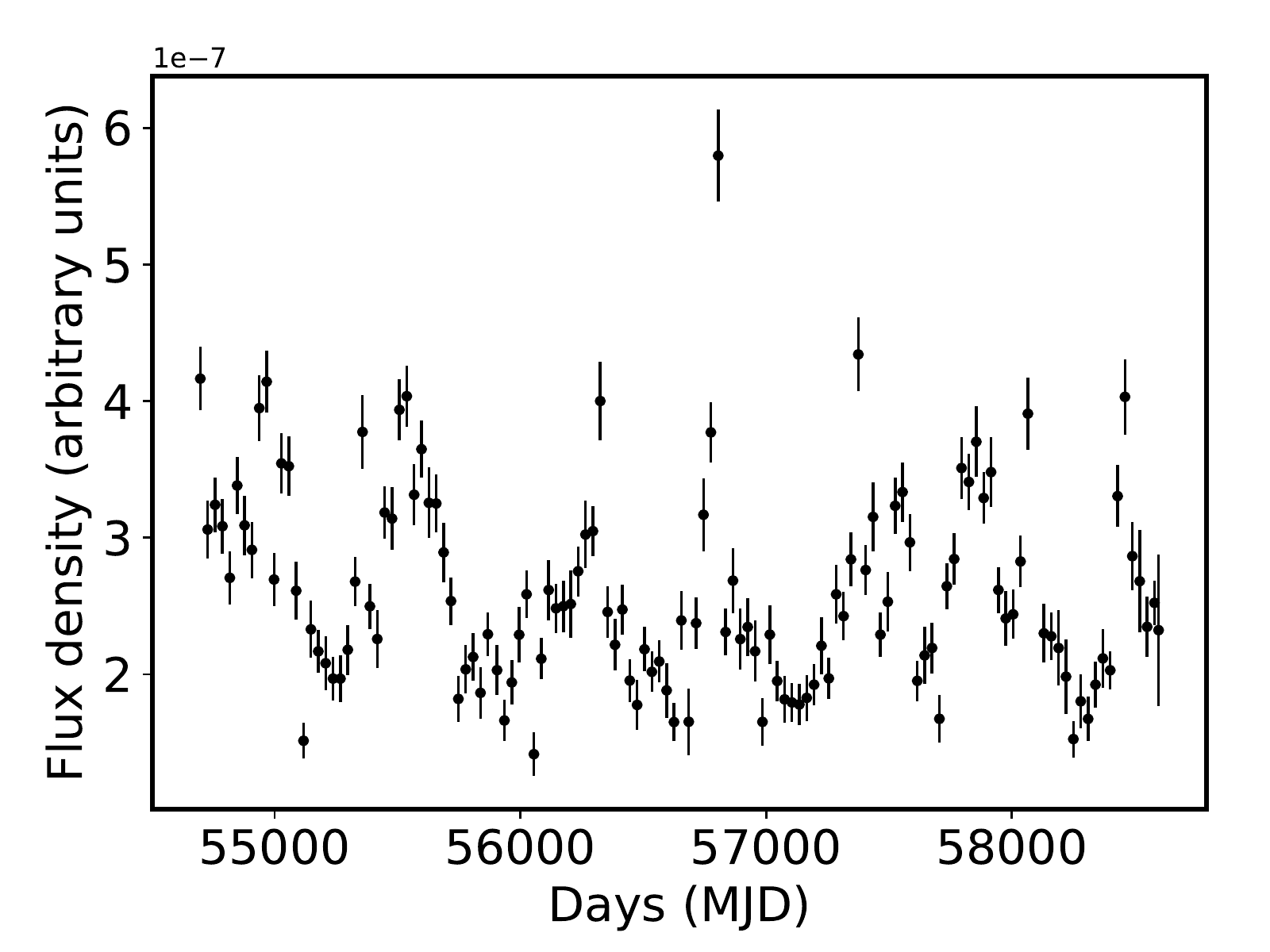} \\
\end{tabular}
\caption{Optical ($R$-band, left column) and high-energy (100\,MeV to 200\,GeV, right column) light curves for the blazars considered in this study. From top to bottom, PG\,1553+113 and PKS\,2155-304, respectively. The optical data are shown here with the original sampling, while for the periodicity analysis they are binned at 30\,days, analogously to the high-energy data.}
\label{fig:lc}
\end{figure*}

We selected two BL\,Lac objects based on the availability of well sampled optical light curves, {\it Fermi} high-energy data, and claims proposed in the recent literature for possible periodicities (Sect.\,\ref{sec:intro}). The sources are \object{PG\,1553+113}, at a redshift $z \gtrsim 0.30$ \citep{Landonietal2014}, and \object{PKS\,2155-304}, at a redshift $z \sim 0.12$ \citep{Falomoetal1991}. 

The optical data ($R$-band) are from several different telescopes and were already discussed in \citet{Sandrinellietal2014,Sandrinellietal2016a,Sandrinellietal2018}. Additional optical data covering the more recent epochs are reported in Appendix\,\ref{sec:newoptdata}. We refer the reader to the quoted papers for all the details about data reduction and analysis. The 100\,MeV to 200\,GeV {\it Fermi} data have been updated with the latest observations discussed in \citet{Covinoetal2019}. Given we are mainly interested in rather long (several months) possible periodicites, optical and high-energy data are binned with a 30\,day sampling. These datasets cover more than a decade of observations for all the sources considered in this work. {\it Fermi} data are regularly sampled, while optical data present gaps due to seasonal visibility or other problems affecting the observations (Fig.\,\ref{fig:lc}).

For PG\,1553+113, a period of $T \sim 796$\,days analyzing the {\it Fermi}/LAT light-curve was proposed by \citet{Ackermannetal2015} and the same period was reported to be consistent with data in other bands \citep[e.g.,][]{Cutinietal2016,Stamerraetal2016}. With a longer coverage the periodicity in the {\it Fermi} data was confirmed by \citet{Tavanietal2018}. A consistent periodicity in the optical data, together with a confirmation at high energies, was discussed by \citet{Sandrinellietal2018}.

For PKS\,2155-304, in the optical, a periodic component at slightly less than one year, superposed on a long-term trend with large-amplitude variations, was proposed by \citet{Zhangetal2014}. The same period ($T \sim 315$\,days) was found by \citet{Sandrinellietal2014}, while a period of approximately two times the optical one ($T \sim 642$\,days) was identified analyzing the {\it Fermi}/LAT light-curve \citep[see also][]{Sandrinellietal2016a}. Confirmation of the periodicity at high energies with a longer coverage by the {\it Fermi} satellite was proposed by \citet{Zhangetal2017a}. A re-analysis of more recent data in the optical and high-energy by \citet{Sandrinellietal2018} confirmed the previous findings. 

These periods for both sources were also identified in the {\it Fermi}/LAT data by the systematic search carried out by \citet{ProkhorovMoraghan2017}, while \citet{Covinoetal2019} and \citet{AitBenkhalietal2019} shed some doubt on the claimed significance of the proposed periodicities at high energy for these two sources (and other blazars observed by {\it Fermi}). \citet{Nilssonetal2018} also did not find any evidence for periodicity in the optical data of both sources (and other blazars well observed in the optical).

\begin{figure*}
\begin{tabular}{cc}
\includegraphics[width=8cm]{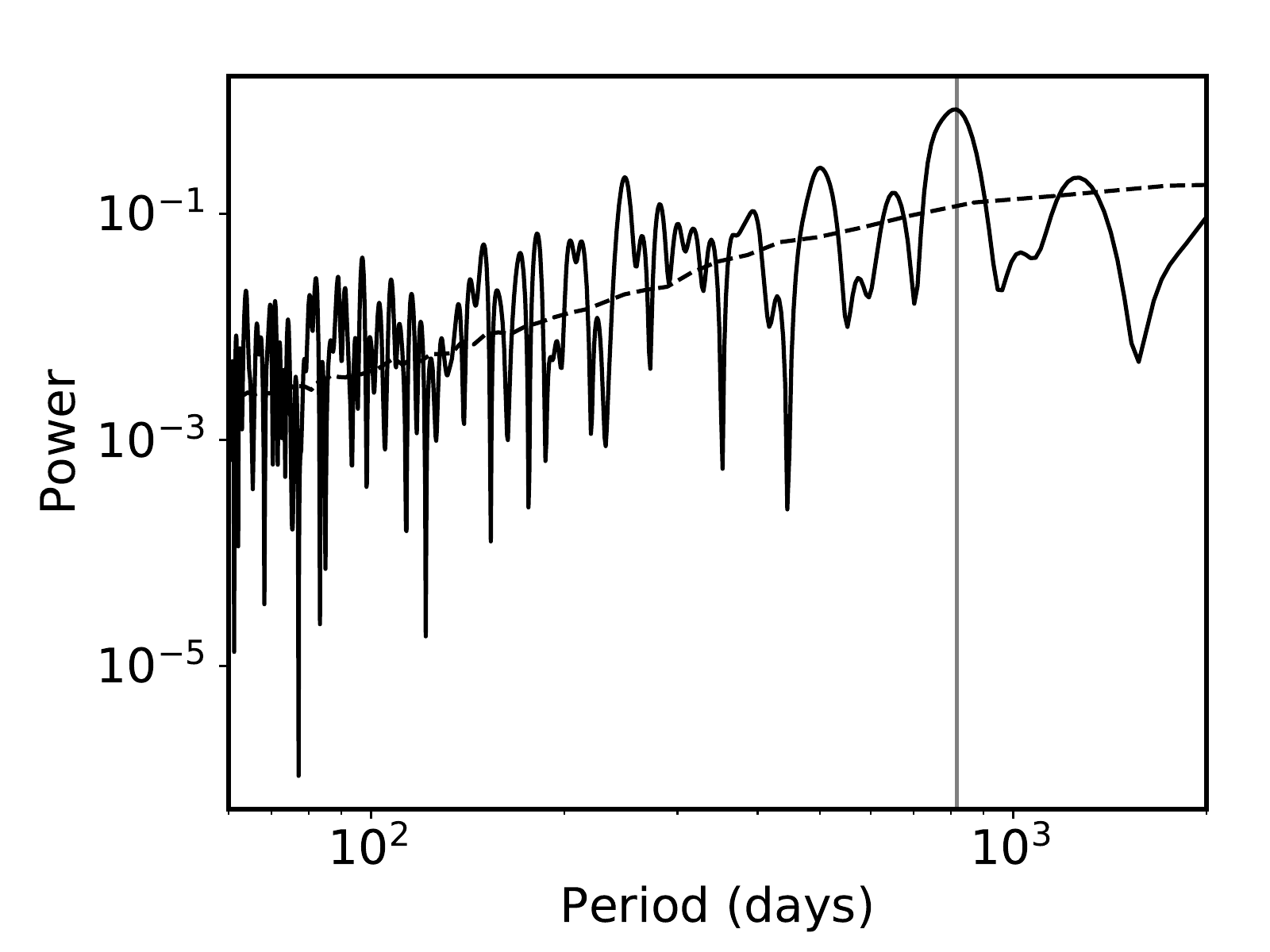} & \includegraphics[width=8cm]{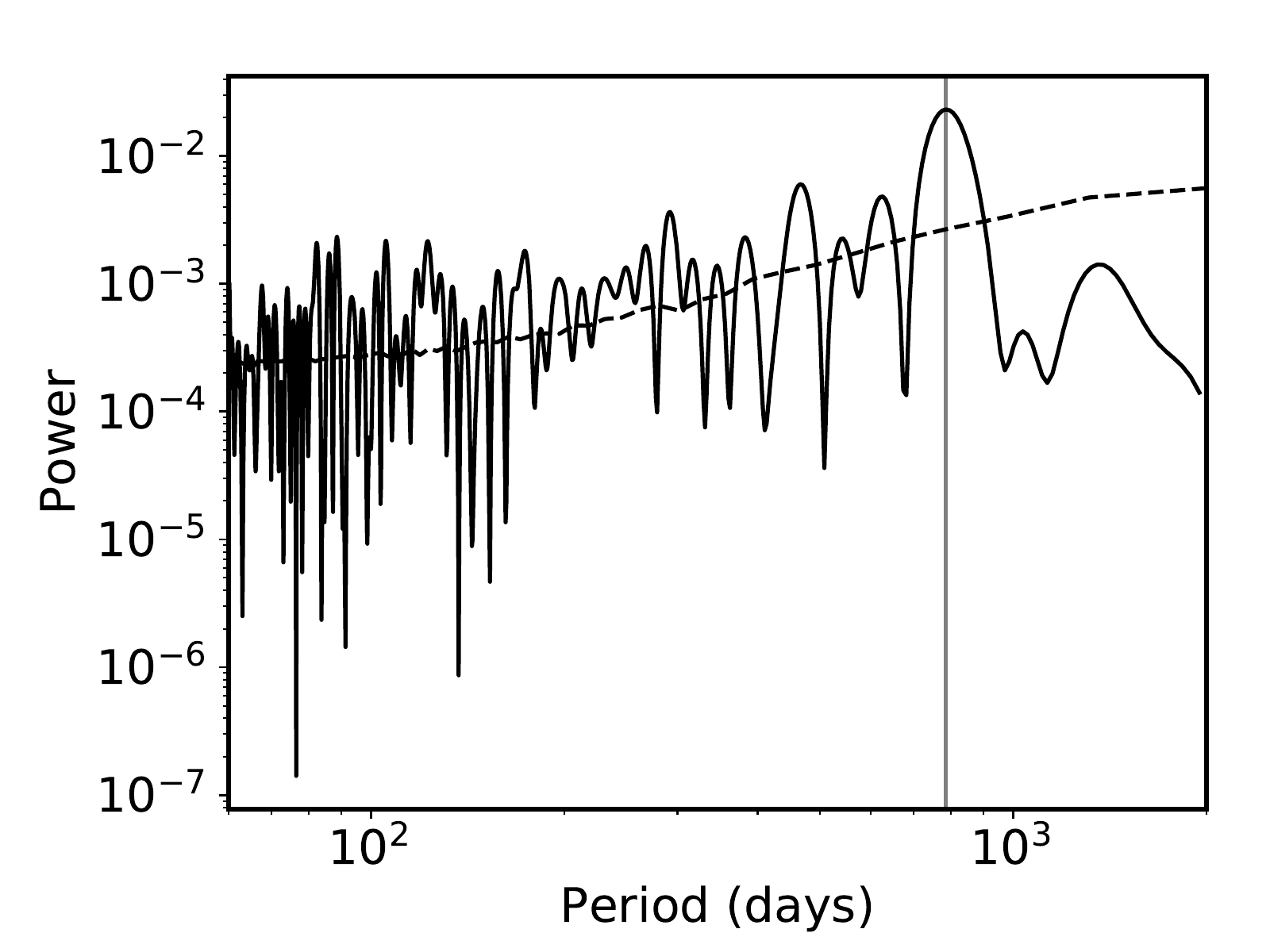} \\
\includegraphics[width=8cm]{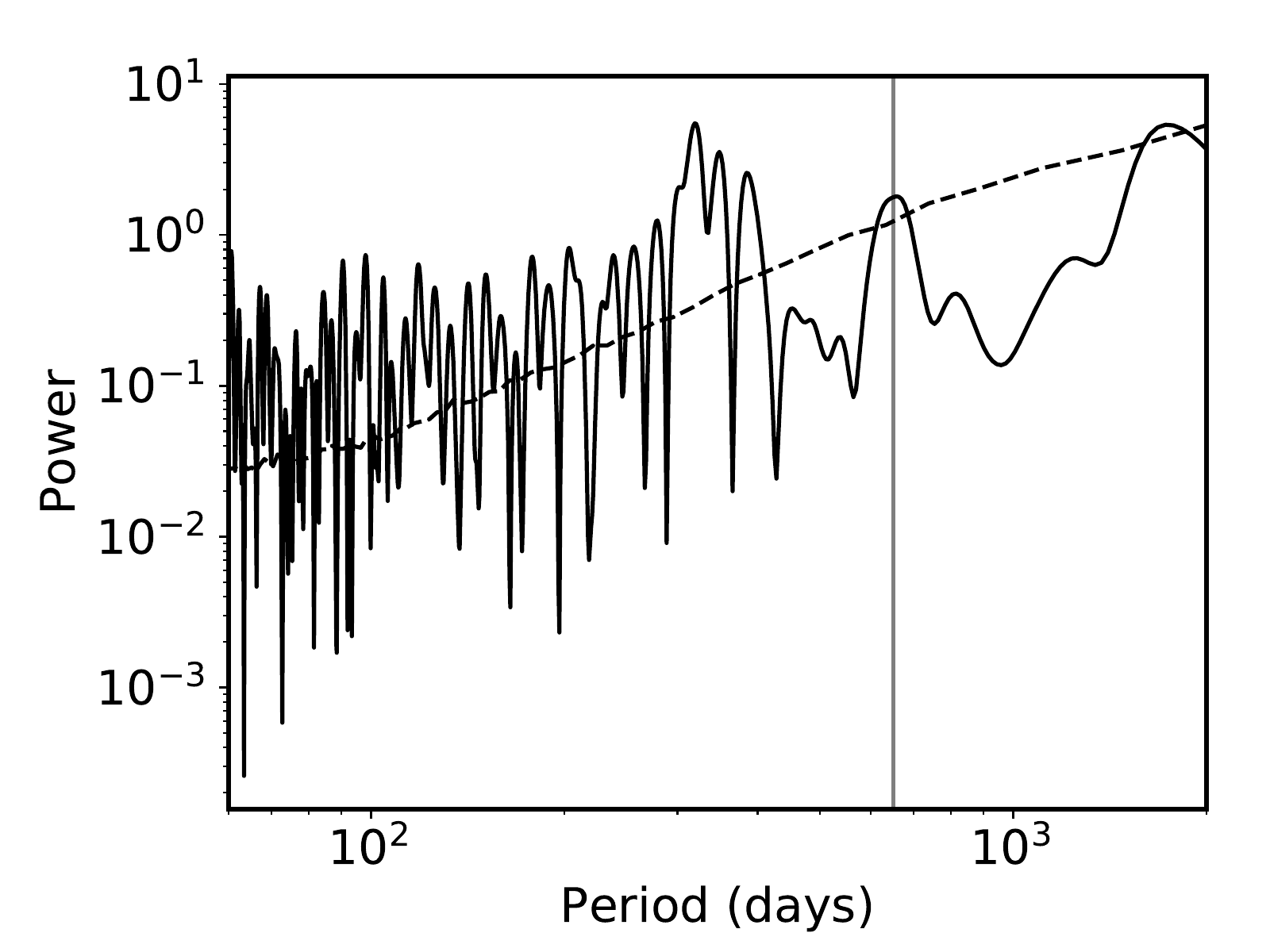} & \includegraphics[width=8cm]{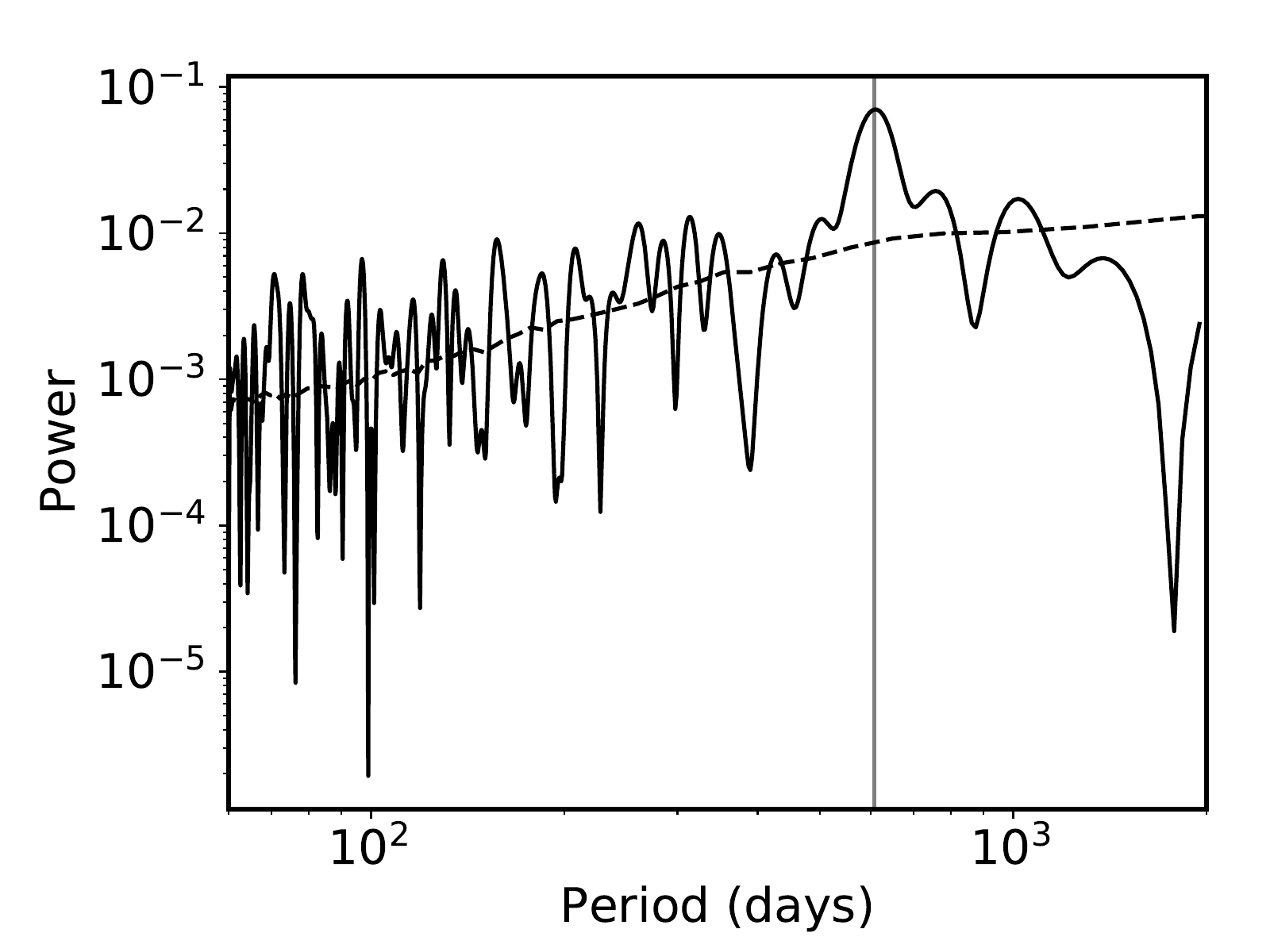} \\
\end{tabular}
\caption{Optical ($R$-band, left column) and high-energy (100\,MeV to 200\,GeV, right column) Lomb-Scargle periodograms based on the  light curves of the blazars considered in this study. From the top to the bottom, PG\,1553+113 and PKS\,2155-304, respectively. The vertical lines indicate the most prominent periods identified in the periodograms (Table\,\ref{tab:LS_periods}). The dashed lines are the PSD for the best fit AE kernel (see Sect(s).\,\ref{sec:meth} and \ref{sec:stat}).}
\label{fig:ls}
\end{figure*}

\section{Results}
\label{sec:res}

\begin{table}
\centering
\begin{tabular}{l|cc}
\hline
\hline
Source & Optical period & High-energy period \\
 & (days) & (days) \\
\hline 
PG\,1553+113 & 820 & 790 \\
PKS\,2155-304 & 650 & 610 \\
\hline
\end{tabular}
\caption{Periods corresponding to the maximum power detected in the Lomb-Scargle periodograms shown in Fig.\,\ref{fig:ls}. The optical and high-energy light curves give consistent periods for PG\,1503+113. For PKS\,2155-304, in the optical, we chose to show a period similar to the high-energy one even if a single dominant feature cannot be easily identified (see Sect.\,\ref{sec:data}).}
\label{tab:LS_periods}
\end{table}

The light curves for the two objects of our interest (Fig.\,\ref{fig:lc}) show intense variability, as it is typical for this category of sources \citep[e.g.,][]{Lindforsetal2016}. The high-energy and optical monitoring both cover more than a decade of observations, therefore allowing us to explore at least a few cycles for year-long periodicities. We first checked whether our light curves are consistent with standard stationarity tests, i.e. the Augmented Dickey-Fuller unit root test (ADF) and the Kwiatkowski-Phillips-Schmidt-Shin (KPSS) test \citep[e.g.,][]{Kwiatkowskietal1992,Hamilton1994}. These tests show that, to some extent, for all light curves but the high-energy data of PKS\,2155-304, non-stationarity is present, as it is easy to infer even after a visual inspection (Fig.\,\ref{fig:lc}) due to the presence of large flares. At variance with other analysis techniques, GP regression however does not typically require one to assume stationarity of the light curves \citep{Kovacevicetal2019} for deriving reliable inferences.

We then carried out a period analysis by means of a generalised LS algorithm \citep{Lomb1976,Scargle1982,Bretthorst2003,VanderPlas2018}. The derived periodograms are shown in Fig.\,\ref{fig:ls} and their maxima are also identified (Table\,\ref{tab:LS_periods}). 
The PKS\,2155-304 LS periodogram (Fig.\,\ref{fig:ls}) does not show a prominent peak, at variance with the other cases here considered. A relatively isolated peak is indeed visible at $\sim 650$\,days, close to the high-energy period. Therefore, following \citet{Chevalieretal2019}, who also considered a periodicity in the optical consistent with the high-energy one, and driven by the intrinsic interest of possible synchronous periodicities in different bands, we report $P \sim 650$\,days in Fig.\,\ref{fig:ls} and Table\,\ref{tab:LS_periods}.

It is also apparent that for both sources the periodograms are characterised by noise increasing toward long periods, the typical behavior when noise is correlated \citep{Press1978,Milotti2002,Milotti2007}. Modeling the noise as power-laws, their indices were previously evaluated in literature \citep[][]{Nilssonetal2018,Covinoetal2019} and they are approximately in the range $1-1.5$. Given that the proposed periodicities are all close to the low-frequency tail of the derived periodograms, any evaluation of their significance has therefore to be carried out with great care \citep[see Sect.\,\ref{sec:tsa} and discussion in][]{Bhatta&Dhital2019}.

\subsection{Stationary kernels}
\label{sec:stat}

For all the light curves considered in this study, we could obtain reasonable fits by GPs with any of the kernel discussed in Sect.\,\ref{sec:meth}. This is not surprising, given the flexibility provided by GP regression, as widely discussed in literature \citep{Ivezicetal2014,Littlefairetal2017,Foreman-Mackeyetal2017,Angusetal2018}. As mentioned in Sect.\,\ref{sec:meth}, the AE and the RQ kernels are however better suited to model our blazar light curves. Together with the considerations discussed in \citet{Wilkins2019}, the choice is also supported by more formal arguments, since both the RQ and the AE kernel functions are  preferred over the simpler SE kernel based upon a Bayesian model comparison. In turn, the AE kernel is only moderately preferred over the RQ kernel, and for the next steps of our analysis we considered both the possibilities. In Table\,\ref{tab:BFserq} we report the computed Bayes factors. The large uninformative priors adopted in the analysis are reported in Appendix\,\ref{ap:priors}. 

If the posterior probabilities of two competing models, e.g. models "0" and "1", are, respectively, $p_{0}$ and $p_{1}$, Bayes factors can be easily converted to probabilities conditioned on the data in favor of model "1" with respect to model "0" as \citep[e.g.,][]{Trotta2007}:
\begin{equation}
p = p_{1}/(p_0+p_1) = BF_{0-1}/(BF_{0-1} + 1),
\end{equation}
where $BF_{0-1} = p_0/p_1$. Therefore $BF \sim 1$ means $p \sim 50$\%, $BF \sim 10$ is slightly larger than $p \sim 90$\% and $BF \sim 100$ gives $p \sim 99$\%. 

\begin{table}
\centering
\begin{tabular}{l|crrr}
\hline
\hline
Source & band & BF$_{SE-RQ}$   & BF$_{SE-AE}$ & BF$_{RQ-AE}$  \\
\hline 
PG\,1553+113 & HE & $14\pm{2}$ & $85 \pm 6$ & $6 \pm 1$ \\
             & Opt & $> 1000$ & $> 1000$ & $4 \pm 1$ \\
PKS\,2155-304 & HE & $71\pm{10}$ & $445 \pm 28$ & $6 \pm 1$\\
             & Opt & $> 1000$ & $> 1000$ & $4 \pm 1$ \\
\hline
\end{tabular}
\caption{Bayes factors for the high-energy (HE) and the optical (Opt) light-curves and the associated probabilities conditioned on the data of supporting one model (e.g. RQ or AE) over another (SE or RQ). Their 1$\sigma$ credible regions are also reported. Flat uninformative or Jeffrey priors on the parameters were added to the likelihood function.}
\label{tab:BFserq}
\end{table}


For stationary kernels, covariance functions and PSDs are Fourier duals \citep{Rasmussen&Williams2006}, and in Fig.\,\ref{fig:ls} we also show the PSD derived by the AE kernel with the best-fit parameters reported in Appendix\,\ref{ap:bestfit}.

The best fit values of the parameters (also often called ``hyper-parameters") of the kernel functions are not in general of straightforward interpretation. The relatively large values for the correlation length parameter $L$, for the AE kernel and the low values of the {\tt gamma} distribution parameter ``$\alpha$" for the RQ kernel (Appendix\,\ref{ap:bestfit}) imply a correlation slowly decaying with increasing data separation \citep[][]{Rasmussen&Williams2006}. This is in qualitative agreement with the long-term noise correlation singled out by modeling periodograms for these sources, as e.g. in \citet{Nilssonetal2018} and \citet{Covinoetal2019}.

\subsection{Periodic kernels}
\label{sec:qp}

Adding a periodic component to the kernel function increases the complexity of the analysis and of course the capability of the model to reproduce the data and the associated noise. In the present work, we leave the kernel parameters essentially unconstrained with uninformative large flat or Jeffrey priors (see Appendix\,\ref{ap:priors}). The periods, in particular, are constrained to be within a large (100-2000\,day) flat range \citep[see, e.g., ][for a discussion about prior role in Bayesian model comparison]{Kass&Raftery1995,Trotta2008}. All the adopted priors are properly normalized in order that the obtained posterior distributions are actual probability distributions \citep[e.g., ][]{Taketal2018}. 

We have explored all the combinations of the AE and RQ kernels with the periodic CS kernel, namely AE$\times$CS, RQ$\times$CS, AE+CS and RQ+CS, listed with increasing number of free parameters (3, 4, 4 and 5, respectively). Covariance functions including a periodic component typically yield similar results, i.e. better posterior probabilities, compared to the stationary kernel description only, as measured by the computed Bayes factors (Table\,\ref{tab:BF}). The preference for the description including a periodicity is partly expected basing on the results of the LS analysis and also because the two sources here considered were selected for the relevance of past periodicity studies. 
The AE$\times$CS combination turns out to be too simple, since the rapid decay of the covariance described by the AE kernel for separations larger than a few times the correlation length make the multiplied periodic kernel essentially ineffective. This is not the case for the RQ kernel since its greater flexibility allows it to model correlation on longer scales. 



\begin{table*}
\centering
\begin{tabular}{l|crccc}
\hline
\hline
Source & band & BF$_{AE-(AE\times CS)}$   & BF$_{RQ-(RQ\times CS)}$ & BF$_{AE-(AE+CS)}$ & BF$_{RQ-(RQ+CS)}$  \\
\hline 
PG\,1553+113 & HE & $2.0\pm{0.2}$ & $202\pm{40}$ & $223\pm{53}$  & $1709\pm{410}$ \\
             & Opt & $7.2\pm{0.7}$ & $261\pm{51}$ & $115\pm{35}$ & $866\pm{222}$ \\ 
PKS\,2155-304 & HE & $1.0\pm{0.1}$ & $50\pm{11}$ & $5\pm{1}$  & $27\pm{12}$ \\
             & Opt & $<< 1$ & $<< 1$ & $1.3\pm{0.2}$ & $1.0\pm{0.2}$ \\
\hline
\end{tabular}
\caption{Bayes factors for the high-energy (HE) and the optical (Opt) light-curves for the various combinations of stationary and period kernels considered in this study. The 1$\sigma$ credible regions for the computed Bayes factors are also reported. Flat uninformative or Jeffrey priors on the parameters were added to the likelihood function.}
\label{tab:BF}
\end{table*}

Results of the stationary vs. period kernel analysis show that there is a rather solid preference for a modeling requiring a period component for PG\,1553+113, both at high energies and in the optical. On the contrary, for PKS\,2155-304, the preference for a periodic component is weaker at high energy and not supported by the data in the optical. 

\begin{figure*}
\begin{tabular}{cc}
\includegraphics[width=8cm]{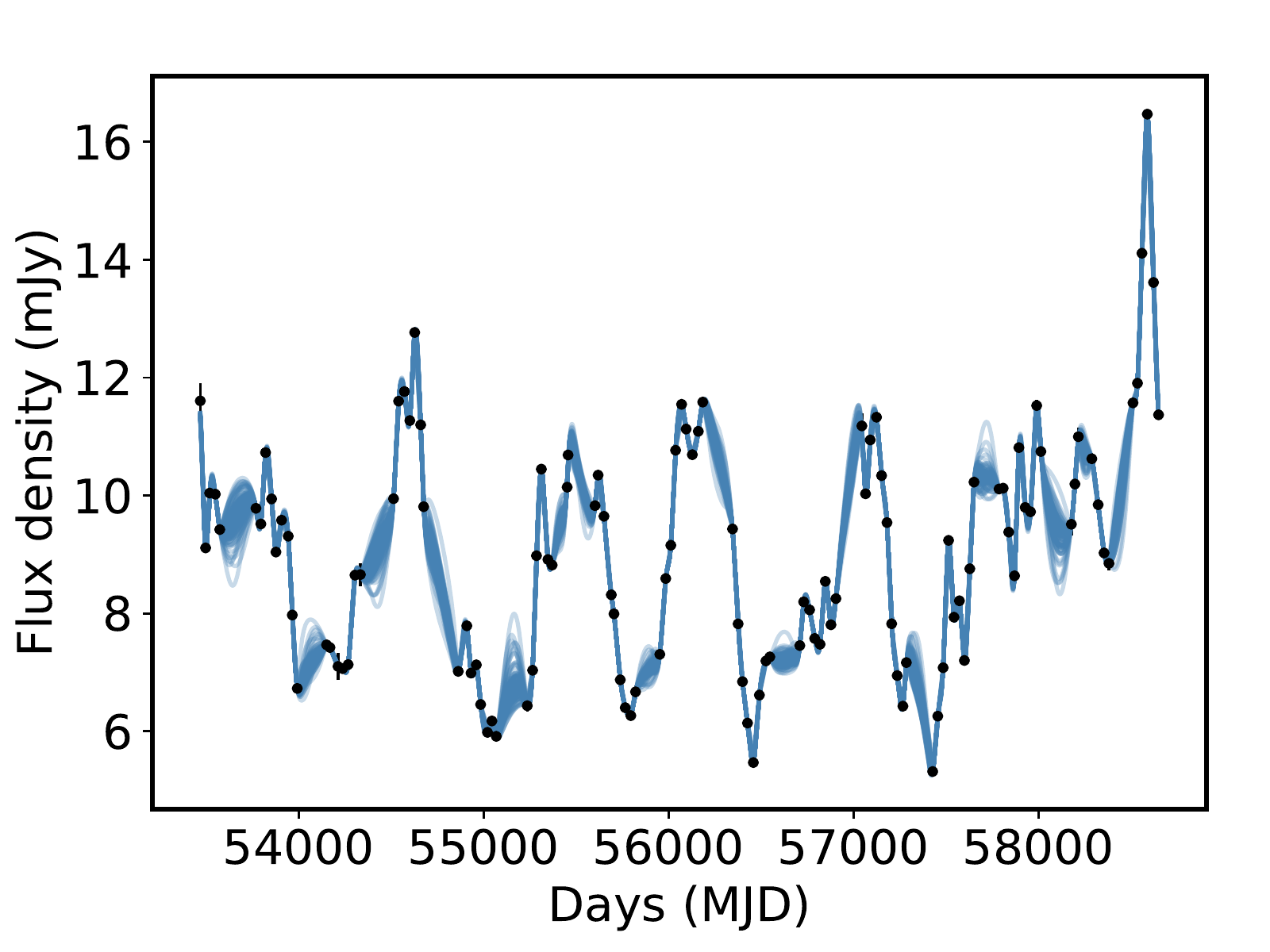} & \includegraphics[width=8cm]{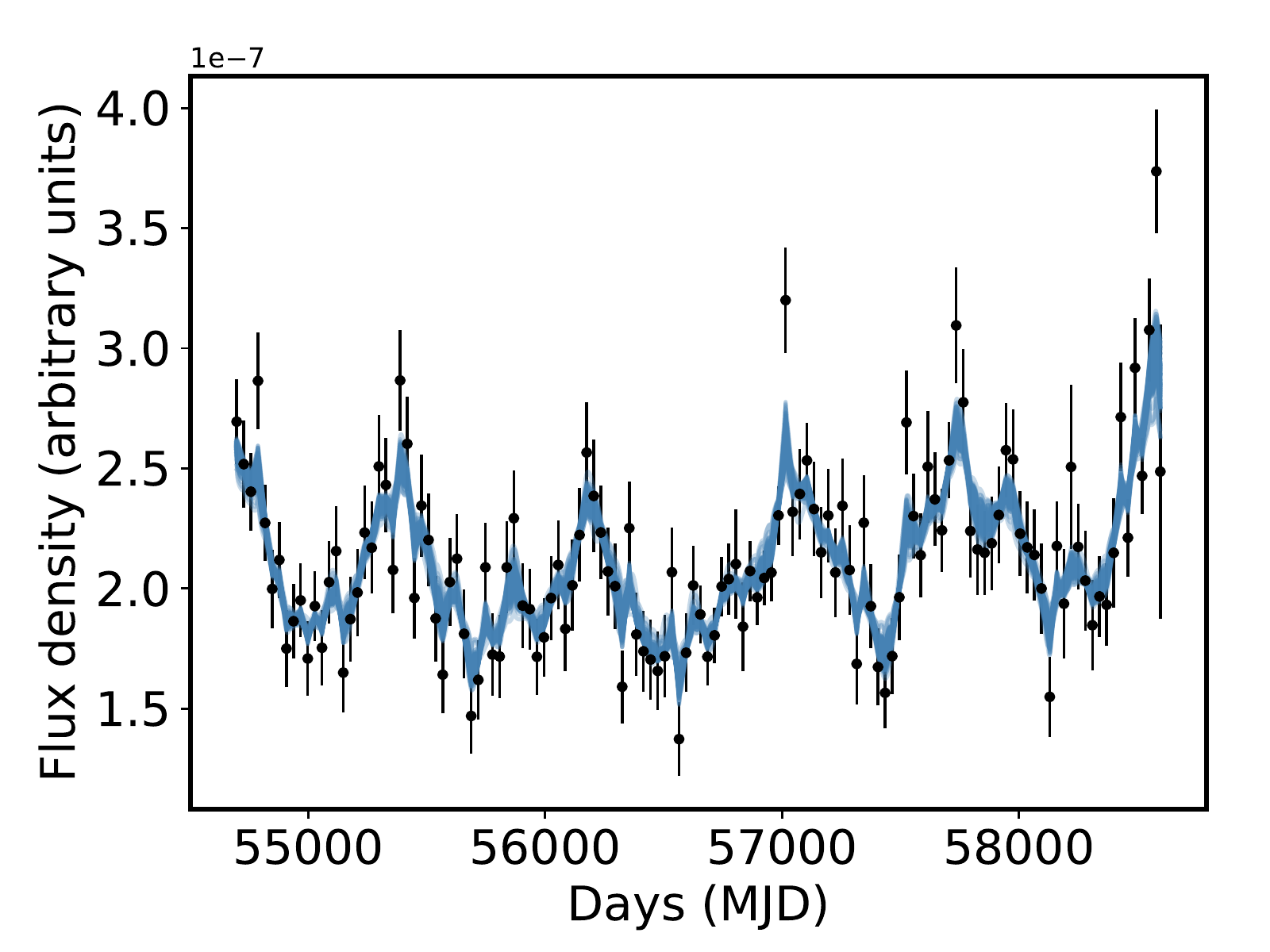} \\
\includegraphics[width=8cm]{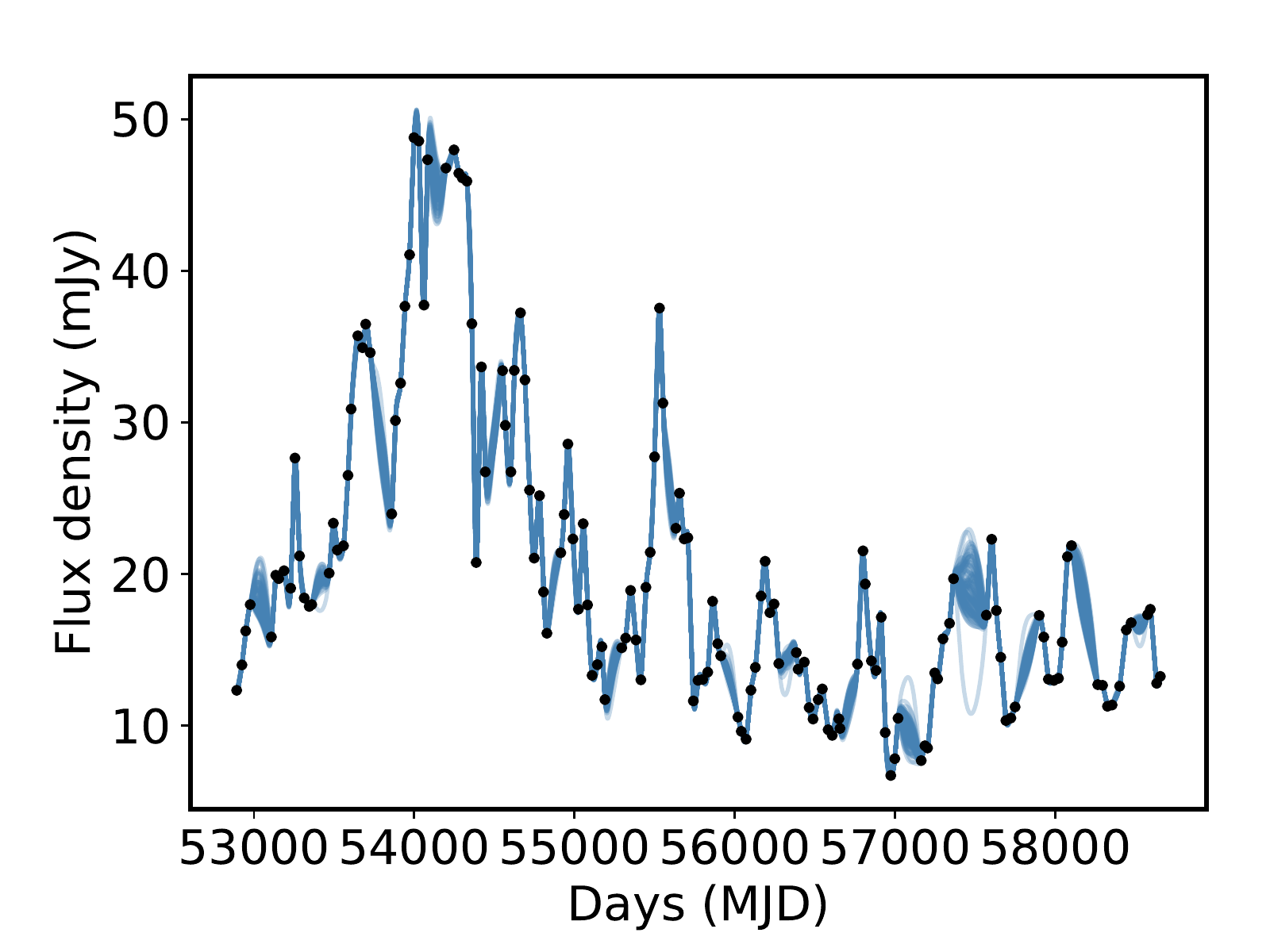} & \includegraphics[width=8cm]{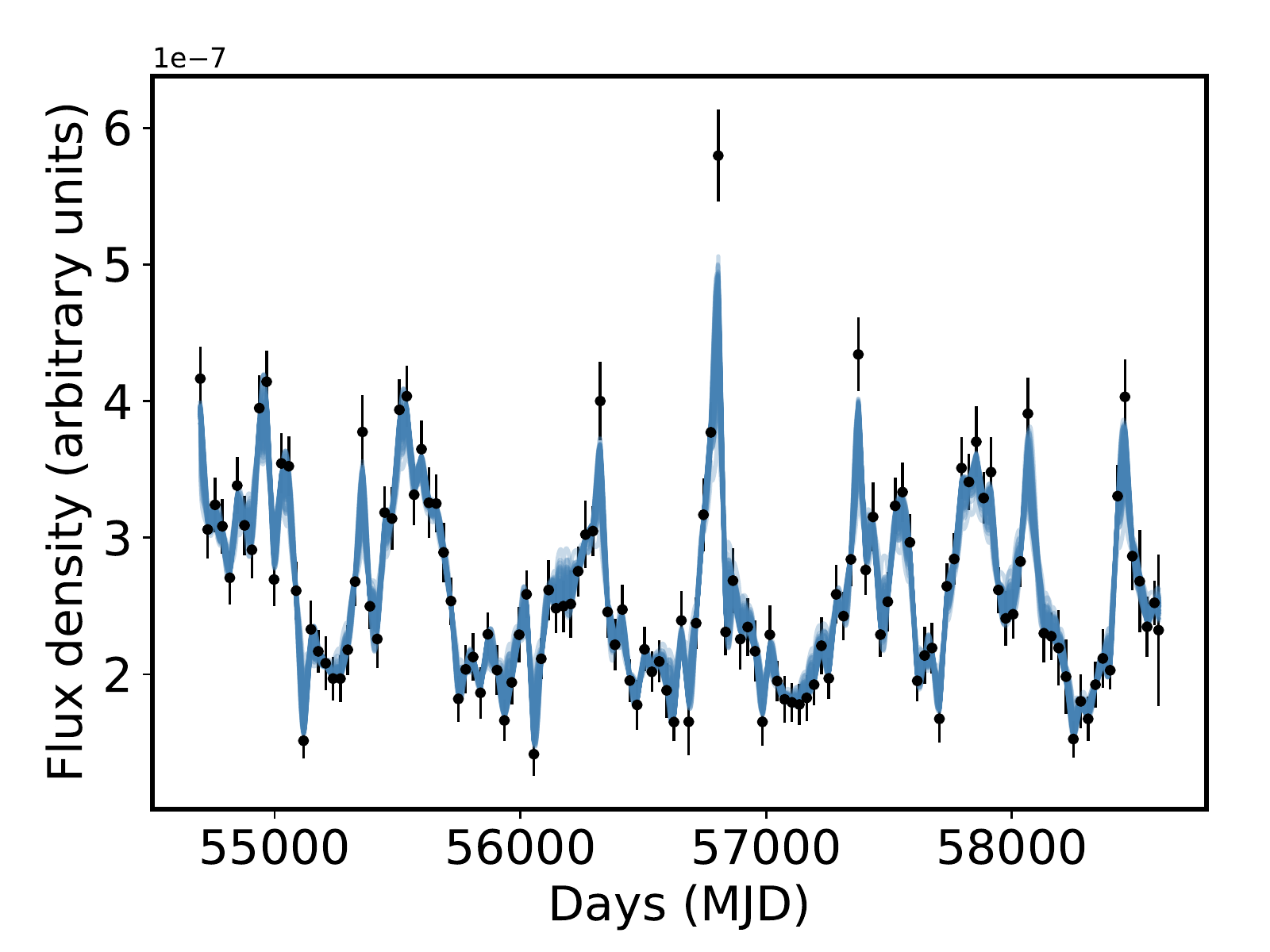} \\
\end{tabular}
\caption{Fit to the light-curve based on a GP regression with a RQ$\times$CS} periodic kernel. Superposed to the light-curve a sample of 100 random set of parameters drawn for the posterior distribution is plot. Optical data ($R$-band, left column) and high-energy data (100\,MeV to 200\,GeV, right column). From the top to the bottom, PG\,1553+113 and PKS\,2155-304, respectively. The optical data are binned with 30\,day sampling analogously to the high-energy data.
\label{fig:fit_qrq}
\end{figure*}

In Fig.\,\ref{fig:fit_qrq} a random selection of possible solutions, with the RQ$\times$CS kernel, extracted for the posterior distribution of the parameters is shown superposed to the original light curves.

\begin{figure*}
\begin{tabular}{cc}
\includegraphics[width=8cm]{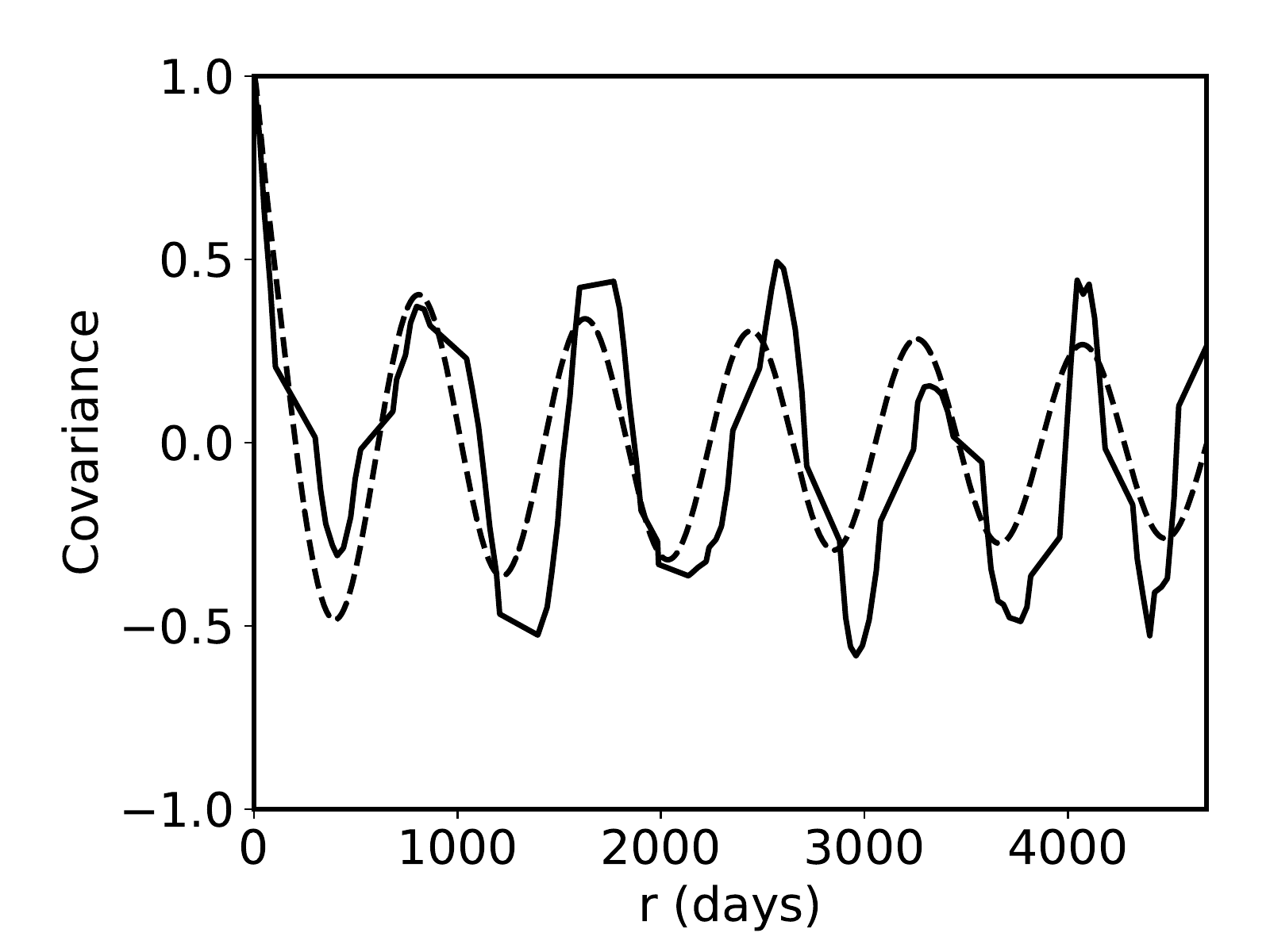} & \includegraphics[width=8cm]{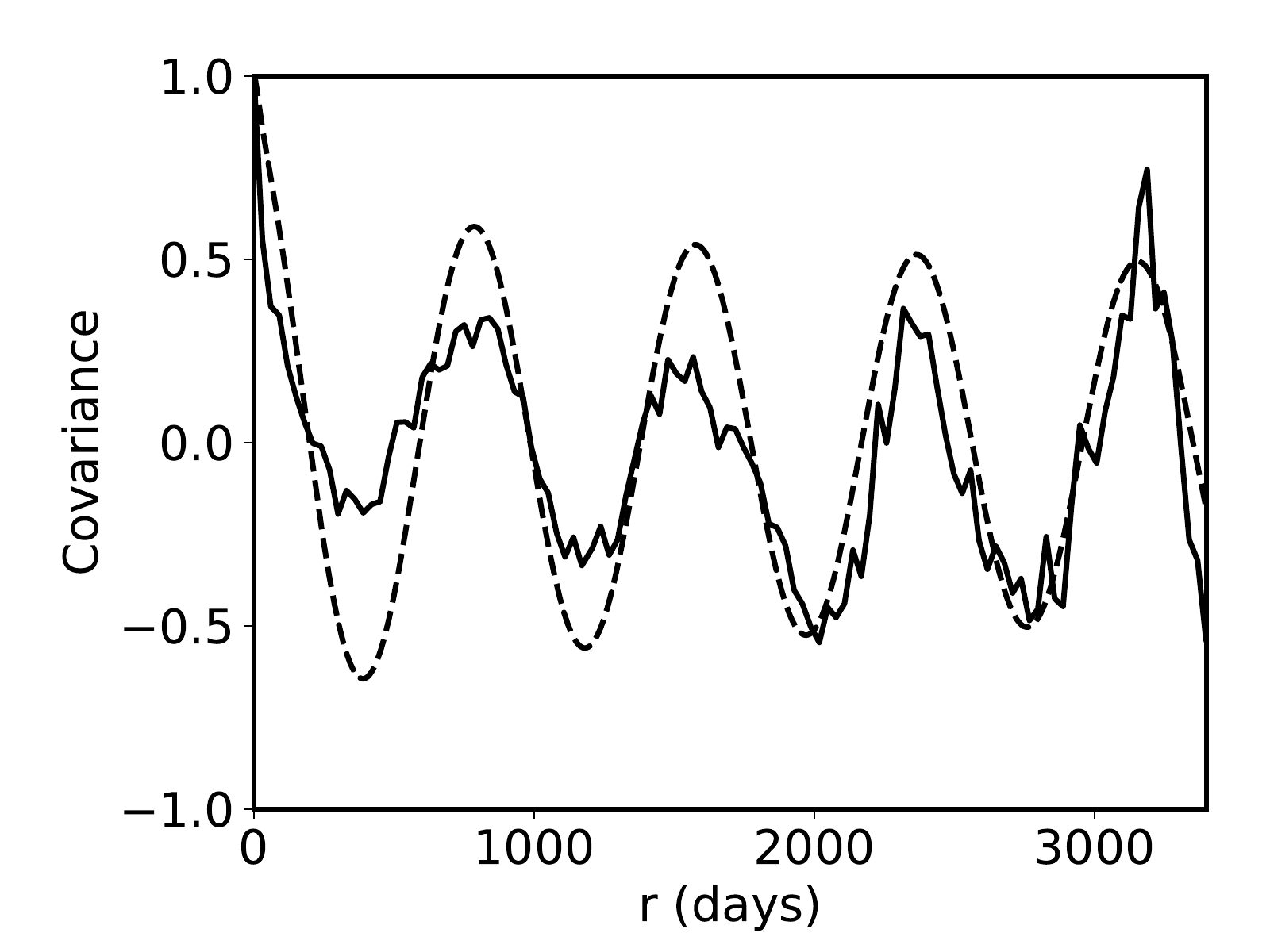} \\
\includegraphics[width=8cm]{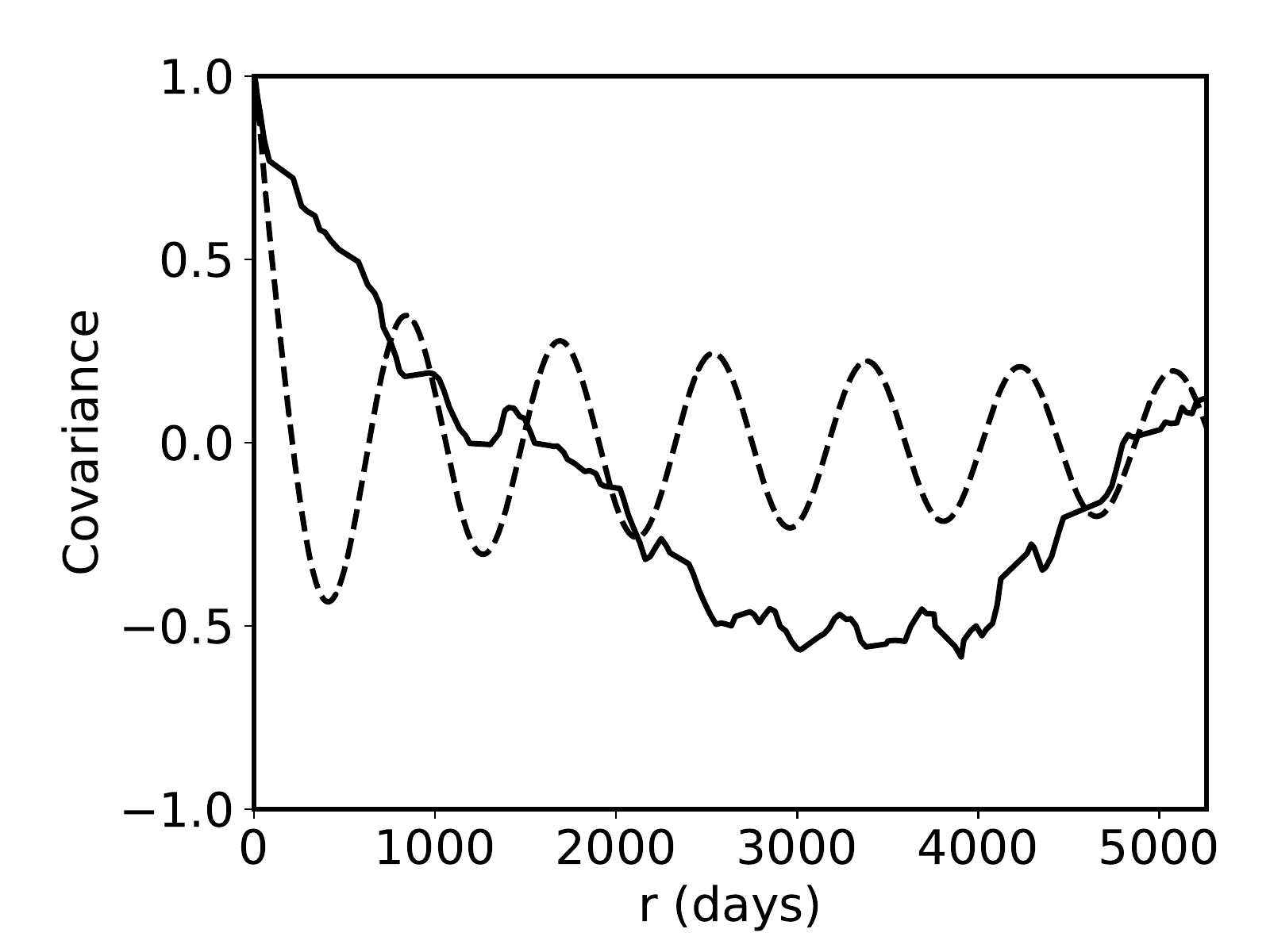} & \includegraphics[width=8cm]{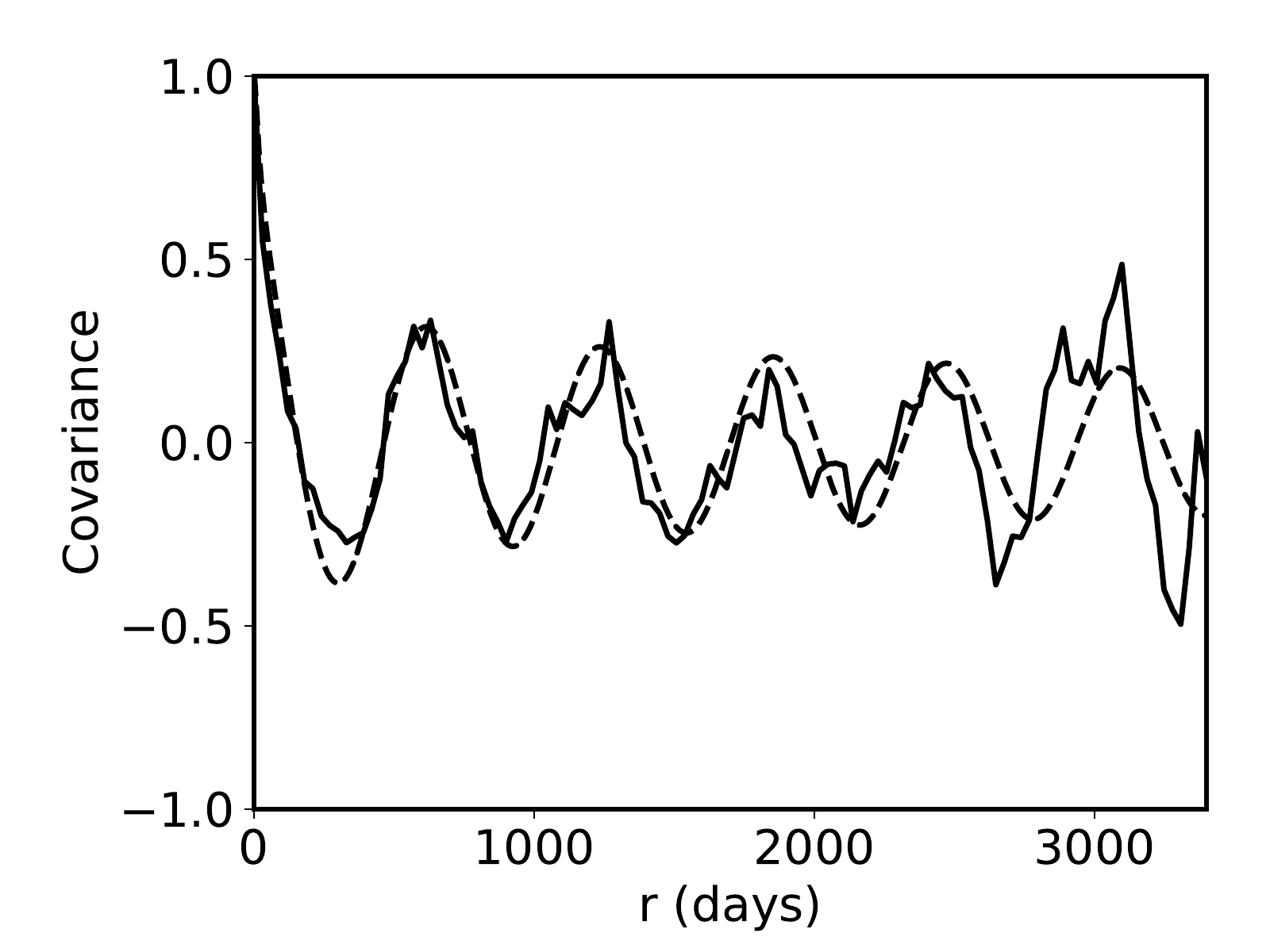} \\
\end{tabular}
\caption{Plot of the best-fit RQ$\times$CS kernel functions together with the ACF computed from the data. The quantity in abscissa, $r = (t_i-t_j)$, is the separation between data points. Optical data ($R$-band, left column) and high-energy data (100\,MeV to 200\,GeV, right column). From the top to the bottom, PG\,1553+113 and PKS\,2155-304, respectively. Dashed line for the best-fit kernel and solid line for the ACF.}
\label{fig:kernel}
\end{figure*}

Even if the addition of a kernel with a periodic component is favoured by the data, it is still worth checking the obtained solutions in order to reach a better insight about the meaning of the GP regression results. To this aim, in Fig.\,\ref{fig:kernel}, we plot the auto-correlation functions (ACF) and the best-fit based on the RQ$\times$CS kernel function. It is  clear that for PG\,1553+113 and for the high-energy data of PKS\,2155-304 the periodic kernel shows correlation peaks at the derived periods and the adopted covariance function correctly describe the ACF computed on the real data. Clearly, reality is richer than our models and peaks in the ACF are never repeating identically, while our models do. However, the average amplitude of the oscillations characterizing the ACFs for PG\,1553+113 is larger than that for PKS\,2155-304, suggesting the main reason for the lower Bayes factor obtained for the latter source. The case of the optical data of PKS\,2155-304 looks different, with no evidence for an actual periodicity in the data.

A different procedure for evaluating the importance of the addition of a periodic component in the analysis is possible when covariance functions obtained by the sum of a stationary and a periodic kernel are applied (in the present study AE+CS or RQ+CS). An estimate of the role of the periodic component can be derived by the analysis of the variance associated to the latter component with respect to the former (the $A_{AE}, A_{RQ}$ and $A_{CS}$ parameters in Appendix\,\ref{ap:bestfit}), and the marginalized posterior distribution of the variance associated to the periodic kernel. First of all, from the various fit results reported in Appendix\,\ref{ap:bestfit}, we see that for the high energy data of PG\,1553+113 the variance associated to the CS kernel is only slightly lower than the one associated to AE or RQ kernels, while for the optical data the periodic kernel describes only a minor fraction of the variance associated to the stationary kernel. In general, however, the model with the RQ kernel provides more constrained parameters although, for PKS\,2155-304, the variance associated to the CS kernel is always badly defined. The variance associated to the CS kernel for PG\,1553+113 is greater than zero with high confidence (better than 99.97\% level) at high energies and at a lower yet high level (98\%) in the optical. For PKS\,2155-304, the null hypothesis of zero variance associated to the periodic kernel cannot be ruled out even at a much lower confidence level.

\begin{figure*}
\begin{tabular}{cc}
\includegraphics[width=8cm]{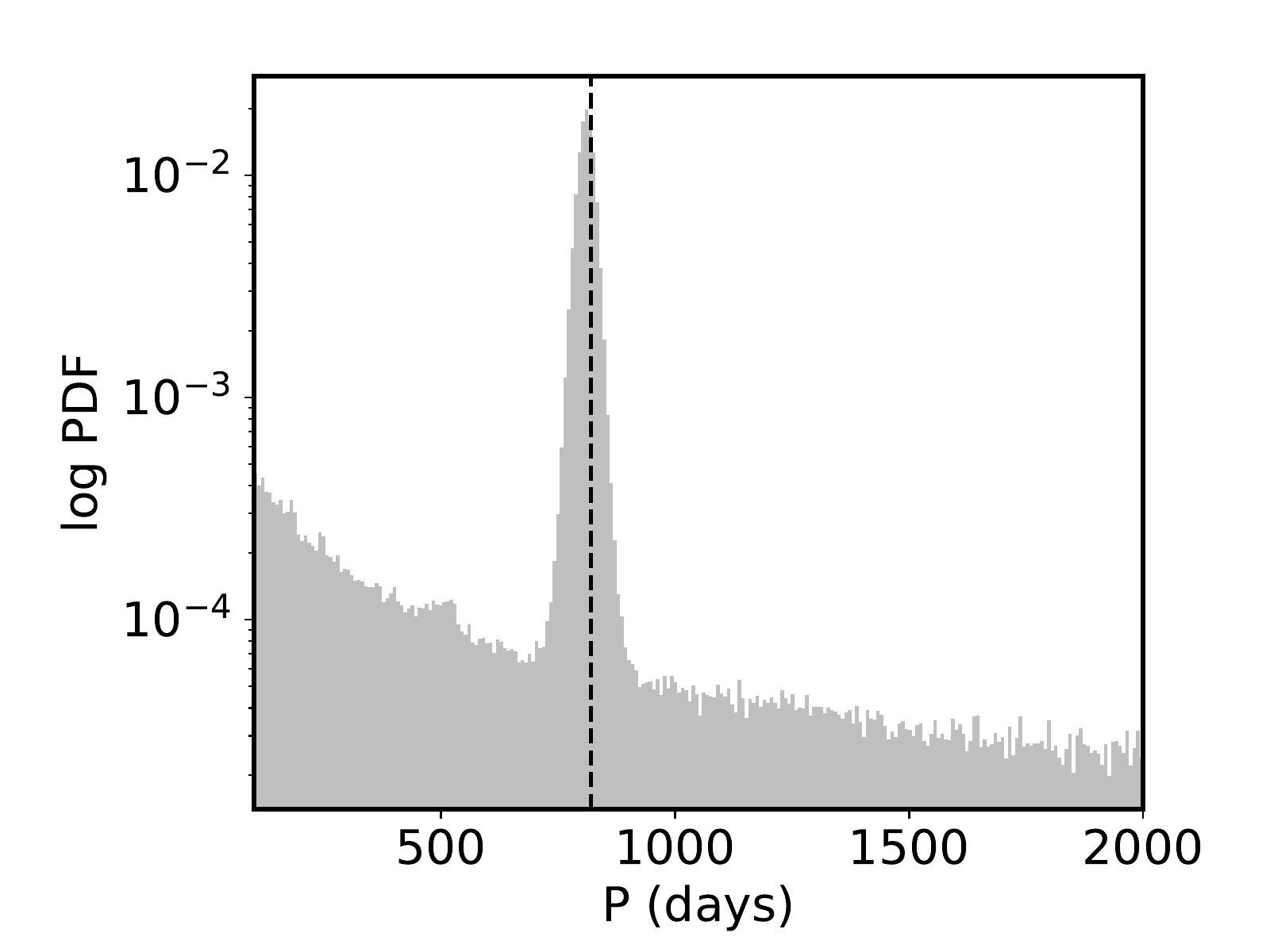} & \includegraphics[width=8cm]{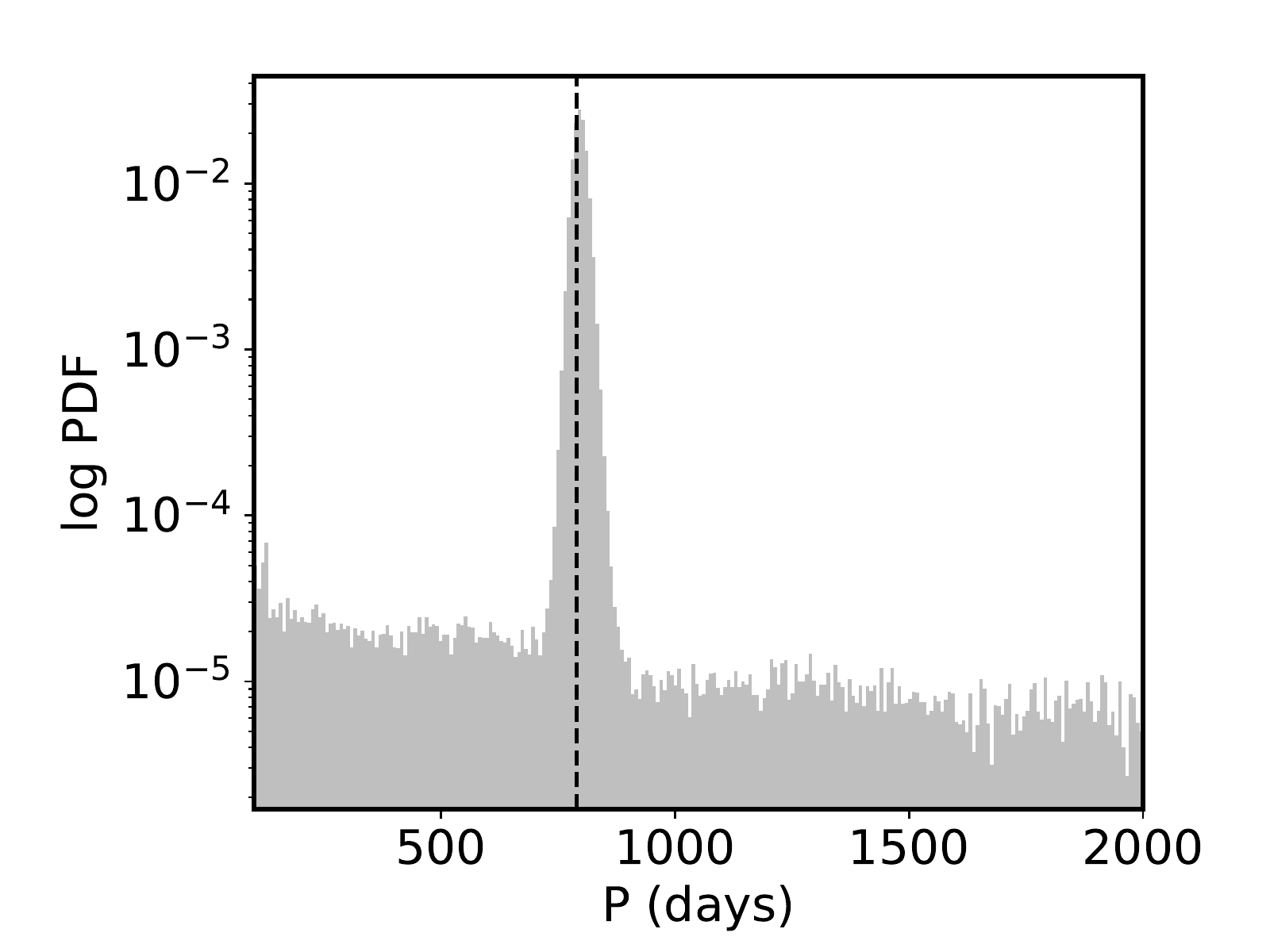} \\
\includegraphics[width=8cm]{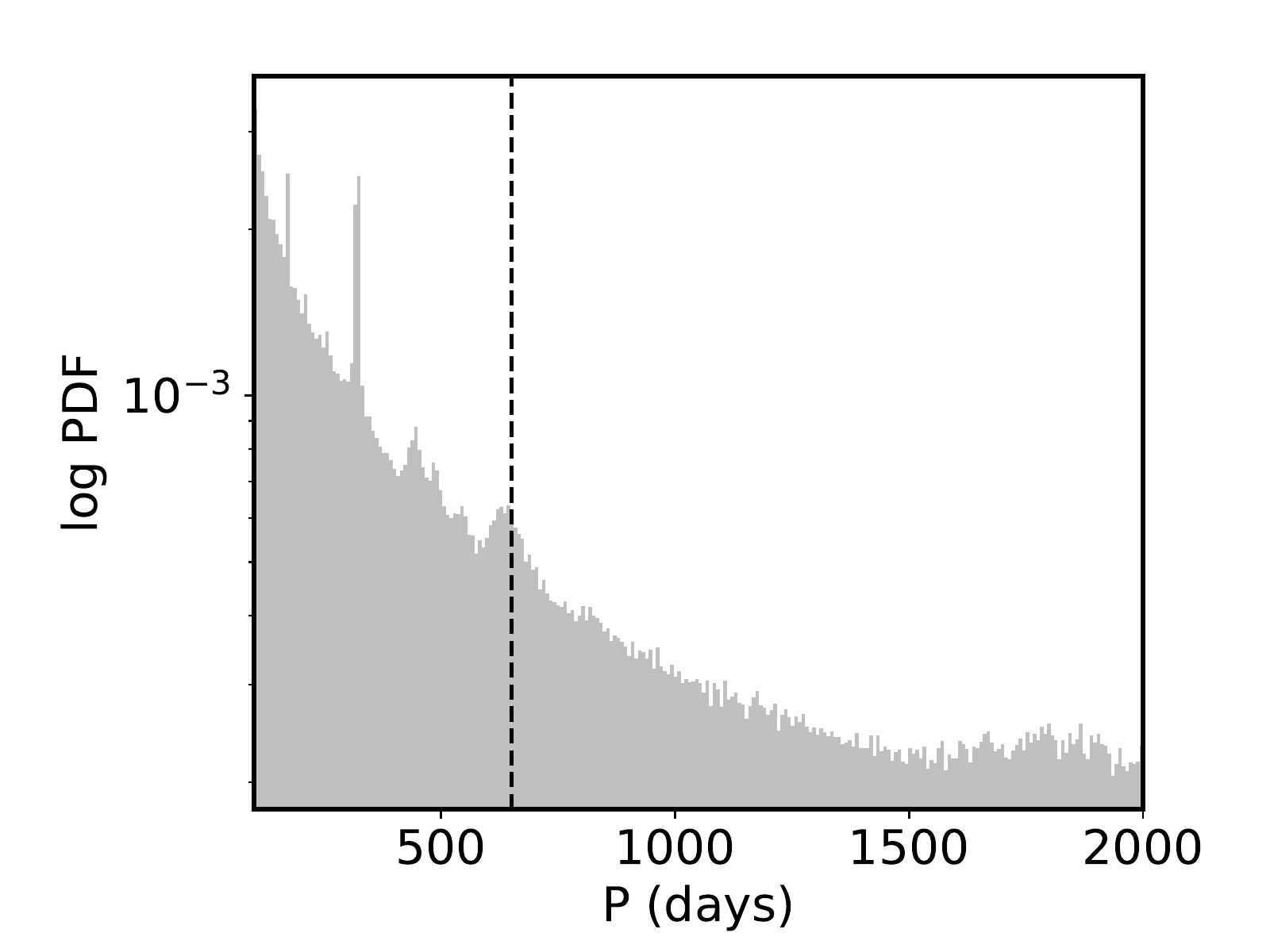} & \includegraphics[width=8cm]{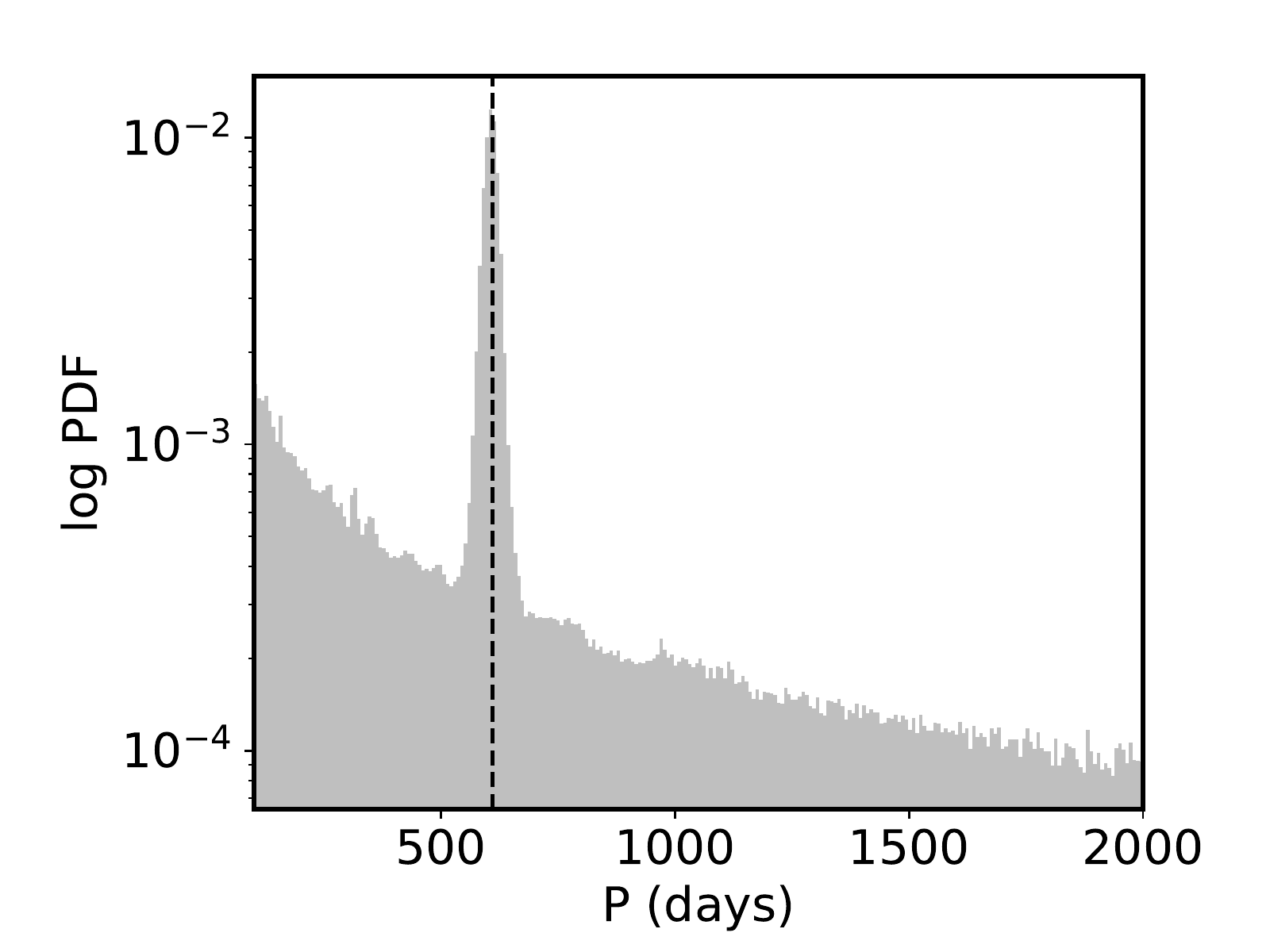} \\
\end{tabular}
\caption{Marginalized posterior probability density functions for the periods singled out with the RQ+CS covariance function (results with other kernel functions are very similar). The PDF are derived by a GP regression and a flat prior from 100 to 2000\,days for the optical data ($R$-band, left column) and high-energy data (100\,MeV to 200\,GeV, right column). From the top to the bottom, PG\,1553+113 and PKS\,2155-304, respectively. The vertical lines indicate the periods identified by the LS analysis (Fig.\,\ref{fig:ls}).}
\label{fig:period_qrq}
\end{figure*}

Finally, we show the marginalized posterior distribution for the periods (Fig.\, \ref{fig:period_qrq}) for the RQ+CS covariance function. This also allows us to investigate on the capabilities of a composite periodic kernel function to identify periodicities. The main peaks identified by the LS analysis (Fig.\,\ref{fig:ls}) are well visible, and the adopted periodicities in the analysis are confirmed. No other periods are singled out, and the distributions are clearly single-peaked with most of the posterior probabilities lying around the identified periods. Again, PKS\,2155-304 in the optical is an exception, and no period stands out for this source, although minor peaks can easily be singled out.

\subsection{Comparison with past results}

We have selected high energy and optical data for two blazar sources with several different claims of possible QPO in literature (see Sect.\,\ref{sec:data}). Summarizing, in our work we find a rather solid evidence supporting the presence of a $\sim 2.2$\,years QPO for PG\,1553+113, while for PKS\,2155-304 the evidence is weaker and still inconclusive at high energies, or there is no evidence at all in the optical. The QPO in PG\,1553+113 was first reported by \citet{Ackermannetal2015} and later confirmed by \citet{Tavanietal2018}, while \citet{Covinoetal2019} and \citet{AitBenkhalietal2019} did not find evidence for it. For PKS\,2155-304, a year-long period in the optical was suggested by \citet{Zhangetal2014} and a hint for a periodicity at high energy, approximately two times longer than the optical one, was suggested by \cite{Sandrinellietal2016a}. No periodicities in the optical for both sources emerged from the analyses by \citet{Nilssonetal2018}. 

It is therefore of interest to see the reasons for these partially (possibly apparently) contradictory results. In general, reporting a possible periodicity always depends on the comparison between a null hypothesis, i.e. the light curve is pure noise, and an hypothesis implying a periodic behavior. Independently of the specific analysis technique applied, any inference is driven by the capability to correctly interpret and model the noise of the data under analysis and by the capability to identify periodic behaviors. These are not trivial tasks and often remain inadequately discussed. The noise is typically modeled, implicitly or explicitly, as PLs or Broken PLs. Then, an excess power at a given frequency is compared to the prediction of the pure noise model. The power can be measured by two general sets of techniques: parametric and non-parametric. The first class includes the Fourier transform \citep[as in][]{Covinoetal2019} or the popular LS algorithm \citep[as in][]{Nilssonetal2018}. These parametric techniques require to model a light curve by a harmonic decomposition, which can be ineffective in identifying even real periods in case of very noisy data and covering only a few cycles of the searched for periodicity \citep[see also discussion in][]{Bhatta&Dhital2019}, as it is the case of the data here considered. On the other hand, non-parametric techniques \citep[as in][and for our GP analysis]{Ackermannetal2015,Tavanietal2018} do not require one to describe the data with a given functional form and might be better suited to extract hidden periodic signals in very noisy data. A full discussion of the pros and cons of parametric and non parametric time-series analysis techniques is certainly beyond the goal of the present study. Yet, we think it is plausible that some of the different claims we mentioned are simply due to difference in the data analysis emphasized by the short coverage, in terms of number of cycles, of the available data.

\section{Conclusions}

In this work we addressed the problem of assessing the significance of periodicities proposed in the literature for two of the most studied blazars: PG\,1553+114 and PKS\,2155-304. We made use of high-energy data, from the {\it Fermi}/LAT instrument, and optical data collected by several telescopes. These data have partly already been analyzed in previous papers (see Sect.\,\ref{sec:data}). The whole topic of blazar year-long possible periodicities is widely discussed in the literature with several papers reporting even contradicting results \citep[see e.g.][for a recent discussion]{Rieger2019}. Our approach is based on GP regression that, in spite of being computationally relatively demanding, offers also several advantages. The analysis does not need to be carried out in the frequency space, with all the possible problems induced by irregular sampling, gaps, etc. The inference relies on a Bayesian model comparison between two possible hypotheses: one only able to describe the correlated noise affecting blazar light curves, and another with in addition a periodic component left free to vary in the range of interest. No assumption about the form of the possible periodic variation was included in the analysis although, in presence of physically motivated scenarios, this could be done. 

Our results, summarized in Tables\,\ref{tab:BFserq} and \ref{tab:BF}, show that the addition of a periodic component always improve the description of the data, in substantial agreement with the various results in literature. For PG\,1553+113 the improvement seems to be relatively solid both at high energies and in the optical, while for PKS\,2155-305 the situation is still inconclusive at high energies and no periodicity seems to be present in the optical.

This is anyway at present mainly an exploratory study. Many aspects of a GP-based analysis need to be more deeply evaluated in future works. One of the most critical topic is about the choice of the covariance functions \citep{Wilkins2019} or the criteria for selecting the best among different possible combinations \citep[e.g.,][]{Duvenaudetal2013}. The problem does not have in general a simple solution. Time series analyses require a modeling of the light curves and of their PSDs. We have here limited our study to the most commonly used kernel functions for GP analysis \citep{Rasmussen&Williams2006}, but the subject is receiving increasing attention in the literature \citep[e.g., ][]{Wilson&PrescottAdams2013,Durrandeetal2016,Foreman-Mackeyetal2017,Tobar2018}.

Another interesting topic is related to the possibility to run a GP regression with multi-dimensional input data, therefore providing a natural environment to study time series available in multiple bands. The analysis can be carried out with different kernel functions in different bands, if needed, and adding a periodic kernel with the same periods for all the input data. This will allow us to optimally use all available information for statistical inference.

\acknowledgments

We thank an anonymous referee for her/his rich and inspiring report. SC is grateful to dr. Luigi Stella and dr. Stefano Andreon for very profitable discussions, and to dr. Ben Granett for having carefully read the manuscript. We acknowledge partial funding from Agenzia Spaziale Italiana-Istituto Nazionale di Astrofisica grant I/004/11/3.

\appendix

\section{Optical data}
\label{sec:newoptdata}

The optical data for PG\,1553+113 and PKS\,2155-304 analysed in this work were partly published in \citet{Sandrinellietal2014,Sandrinellietal2016a,Sandrinellietal2018}. Here we report data obtained after the publication to the present epoch. We refer to the quoted publications for all the details about data reduction and analysis.

Tables\,\ref{tab:1553opt} and \ref{tab:2155opt}, contain data for PG\,1553+113 and PKS\,2155-304, respectively. For both tables, Column\,1 lists the MJD of the observation, Column\,2 gives the $R_c$ magnitude and Column\,3 its 1\,$\sigma$ error.

\begin{deluxetable}{ccc}
\tablecaption{Optical data for PG\,1553+113. Column\,1 lists the MJD of the observation, Column\,2 gives the $R_c$ magnitude and Column\,3 its 1\,$\sigma$ error. \label{tab:1553opt}}
\tablehead{\colhead{MJD} & \colhead{Magnitudes} & \colhead{1\,$\sigma$ error} \\ 
\colhead{(days)} & \colhead{($R_c$ band)} & \colhead{} } 
\startdata
53468.40516 & 13.57 & 0.03 \\
53495.11636 & 13.76 & 0.03 \\
53499.10951 & 13.88 & 0.01 \\
53501.12239 & 13.88 & 0.01 \\
53503.10500 & 13.82 & 0.01 \\
53506.13048 & 13.80 & 0.01 \\
53511.17848 & 13.81 & 0.01 \\
53518.05972 & 13.75 & 0.01 \\
53523.03968 & 13.68 & 0.02 \\
53526.03937 & 13.74 & 0.02 \\
53528.03765 & 13.81 & 0.03 \\
53538.13407 & 13.76 & 0.00 \\
53556.10483 & 13.71 & 0.00 \\
53573.03119 & 13.78 & 0.01 \\
53970.04372 & 14.00 & 0.01 \\
53995.98658 & 14.16 & 0.01 \\
54218.26792 & 14.14 & 0.05 \\
54234.33117 & 14.09 & 0.05 \\
54255.12331 & 14.22 & 0.03 \\
54331.02096 & 13.88 & 0.03 \\
54574.25162 & 13.59 & 0.05 \\
54588.38112 & 13.63 & 0.10 \\
54906.38731 & 13.94 & 0.07 \\
54921.35537 & 14.09 & 0.03 \\
54935.38222 & 14.13 & 0.06 \\
54950.31922 & 14.08 & 0.03 \\
54966.27467 & 14.04 & 0.03 \\
54982.33886 & 14.23 & 0.03 \\
55005.27961 & 13.95 & 0.05 \\
55028.21905 & 14.31 & 0.03 \\
55982.37299 & 13.97 & 0.07 \\
56011.21562 & 13.89 & 0.02 \\
56013.24880 & 13.87 & 0.02 \\
56016.39511 & 13.79 & 0.02 \\
56019.32278 & 13.79 & 0.01 \\
56021.33930 & 13.76 & 0.01 \\
56026.22237 & 13.82 & 0.03 \\
56029.23440 & 13.83 & 0.01 \\
56034.21658 & 13.76 & 0.02 \\
56036.34757 & 13.63 & 0.02 \\
56042.22173 & 13.63 & 0.02 \\
56047.22496 & 13.59 & 0.03 \\
56062.35472 & 13.53 & 0.03 \\
56065.18069 & 13.66 & 0.02 \\
56067.36846 & 13.68 & 0.02 \\
\enddata
\end{deluxetable}

\begin{deluxetable}{ccc}
\tablecaption{Optical data for PKS\,2155-304. Column\,1 lists the MJD of the observation, Column\,2 gives the $R_c$ magnitude and Column\,3 its 1\,$\sigma$ error. \label{tab:2155opt}}
\tablehead{\colhead{MJD} & \colhead{Magnitude} & \colhead{1\,$\sigma$ error} \\ 
\colhead{(days)} & \colhead{($R_c$ band)} & \colhead{} } 
\startdata
55516.04326 & 12.36 & 0.04 \\
55518.11836 & 12.51 & 0.05 \\
55521.02647 & 12.30 & 0.05 \\
55523.04867 & 12.20 & 0.03 \\
55525.14933 & 12.26 & 0.04 \\
55531.10888 & 12.41 & 0.04 \\
55533.13191 & 12.23 & 0.04 \\
55536.13616 & 12.11 & 0.04 \\
55538.11141 & 12.22 & 0.04 \\
55540.08721 & 12.44 & 0.04 \\
55542.08486 & 12.41 & 0.04 \\
55544.06030 & 12.38 & 0.04 \\
55547.10680 & 12.51 & 0.05 \\
55552.04550 & 12.59 & 0.05 \\
55559.08028 & 12.77 & 0.04 \\
55562.06174 & 12.70 & 0.04 \\
55565.04775 & 12.82 & 0.04 \\
55567.06705 & 13.02 & 0.03 \\
55571.05565 & 12.99 & 0.04 \\
\enddata
\end{deluxetable}

We  refer the reader to the reported references for anything related to data reduction, analysis and calibration.

\section{Software packages}
\label{ap:soft}

We have developed software tools and used third-party libraries all developed with the {\tt python} language \citep{vanRossum1995} (v. 3.7)\footnote{http://www.python.org} with the usual set of scientific libraries ({\tt numpy} \citep{numpy} (v. 1.15.4)\footnote{http://www.numpy.org} and {\tt scipy} \citep{scipy} (1.10)\footnote{https://www.scipy.org}. The ADF and KPSS stationariety tests are coded in the {\tt statsmodels} library (v. 0.9.0)\footnote{https://www.statsmodels.org/stable/index.html}. The generalised LS algorithm we applied is part of the {\tt astropy} (v. 3.1.2)\footnote{http://www.astropy.org} suite \citep{astropy2013,astropy2018}. ACF are computed by {\tt numpy} tools. Non-linear optimization algorithms and numerical integration tools are provided by the {\tt minimize} and {\tt integrate} subpackages of {\tt scipy} library. MCMC algorithms are provided by the {\tt emcee}\footnote{http://dfm.io/emcee/current/} (v. 2.2.1) library \citep{Foreman-Mackeyetal2013}. GP analysis is carried out by the {\tt george} package (v. 0.3.1)\footnote{https://george.readthedocs.io/en/latest/} \citep{Ambikasaranetal2014}.
Plots are produced within the {\tt matplotlib} \citep{Hunter2007} (v. 3.0.2)\footnote{https://www.matplotlib.org} framework. Multidimensional projection plots were obtained with the {\tt corner} \citep{Foreman-Mackey2016} (v. 2.0.2)\footnote{https://corner.readthedocs.io/en/latest/} library.

\section{Priors imposed to the analysis}
\label{ap:priors}

Throughout this paper we ave always adopted larger uninformative or Jeffrey priors (Table\,\ref{tab:priors}).

\begin{table}
\centering
\begin{tabular}{c|c}
\hline
\hline
Hyper-parameter & prior \\
\hline 
$\ln A$ & {\it Uniform} [-20, 20] \\
$\ln L$ & {\it Uniform} [-20, 20] \\
$\ln \alpha$ & {\it Uniform} [-10, 10] \\
$P$ & {\it Uniform} [100, 2000] \\
\hline
\end{tabular}
\caption{Prior information adopted for for analyses described in Sect.\,\ref{sec:res}. $P_0$ is the period singled out by the LS analysis. All the priors are properly normalized for the computation of the Bayes factors.}
\label{tab:priors}
\end{table}

\section{Best fit hyper-parameters}
\label{ap:bestfit}

We report here the best fit hyper-parameters obtained fitting our data with the SE kernel (Table\,\ref{tab:SE_hyp}), AE kernel (Table\,\ref{tab:AE_hyp}), the RQ kernel (Table\,\ref{tab:RQ_hyp}), the AE$\times$CS kernel (Table\,\ref{tab:AEC_hyp}), the RQ$\times$CS kernel (Table\,\ref{tab:CRQ_hyp}), the AE+CS kernel (Table\,\ref{tab:CEXs_hyp}) and the RQ+CS kernel (Table\,\ref{tab:CRQs_hyp}).
For our analyses, we have multiplied the high energy and the optical data by factors $10^7$ and $10^3$, respectively, for better numerical optimization. This has of course no effect on the reported results.

\begin{table}
\centering
\begin{tabular}{l|ccc}
\hline
\hline
Source & band & $\ln A$   & $\ln L$ \\
\hline 
PG\,1553+113 & HE & $-2.46^{+0.29}_{-0.26}$ & $4.44^{+0.25}_{-0.39}$\\
             & Opt & $1.17^{+0.14}_{-0.14}$ & $3.42^{+0.04}_{-0.04}$  \\
PKS\,2155-304 & HE & $-0.85^{+0.15}_{-0.14}$ & $4.31^{+0.07}_{-0.07}$\\
             & Opt & $4.36^{+0.13}_{-0.12}$ & $3.46^{+0.03}_{-0.03}$ \\
\hline
\end{tabular}
\caption{1$\sigma$ credible regions and the the maximum a posterior estimator for the hyper-parameters derived by the analysis adopting the SE kernel for the high-energy (HE) and the optical (Opt) light-curves. Flat uninformative or Jeffrey priors on the parameters were added to the likelihood function.}
\label{tab:SE_hyp}
\end{table}

\begin{table}
\centering
\begin{tabular}{l|ccc}
\hline
\hline
Source & band & $\ln A$   & $\ln L$ \\
\hline 
PG\,1553+113 & HE & $-2.14^{+0.51}_{-0.32}$ & $5.24^{+0.65}_{-0.43}$\\
             & Opt & $1.41^{+0.32}_{-0.24}$ & $5.26^{+0.36}_{-0.27}$  \\
PKS\,2155-304 & HE & $-0.73^{+0.22}_{-0.19}$ & $4.31^{+0.28}_{-0.24}$\\
             & Opt & $4.68^{+0.42}_{-0.29}$ &  $5.82^{+0.44}_{-0.31}$ \\
\hline
\end{tabular}
\caption{1$\sigma$ credible regions and the the maximum a posterior estimator for the hyper-parameters derived by the analysis adopting the AE kernel for the high-energy (HE) and the optical (Opt) light-curves. Flat uninformative or Jeffrey priors on the parameters were added to the likelihood function.}
\label{tab:AE_hyp}
\end{table}

\begin{table}
\centering
\begin{tabular}{l|cccc}
\hline
\hline
Source & band & $\ln A$   & $\ln \alpha$  & $\ln L$ \\
\hline 
PG\,1553+113 & HE & $-1.99^{+1.41}_{-0.47}$ & $-1.45^{+1.36}_{-1.97}$               & $4.11^{+0.64}_{-0.41}$\\
             & Opt & $1.83^{+1.45}_{-0.48}$ & $-1.85^{+0.85}_{-1.78}$ & $3.97^{+0.70}_{-0.25}$  \\
PKS\,2155-304 & HE & $-0.63^{+0.72}_{-0.24}$ & $-1.00^{+0.75}_{-1.41}$ & $3.28^{+0.25}_{-0.20}$\\
             & Opt & $6.00^{+1.89}_{-1.04}$ & $-3.92^{+1.19}_{-1.95}$ & $4.53^{+0.94}_{-0.51}$ \\
\hline
\end{tabular}
\caption{1$\sigma$ credible regions and the the maximum a posterior estimator for the hyper-parameters derived by the analysis adopting the RQ kernel for the high-energy (HE) and the optical (Opt) light-curves. Flat uninformative or Jeffrey priors on the parameters were added to the likelihood function.}
\label{tab:RQ_hyp}
\end{table}

\begin{table}
\centering
\begin{tabular}{l|cccc}
\hline
\hline
Source & band & $\ln A$   & $\ln L$ & $\ln P$ \\
\hline 
PG\,1553+113 & HE & $-2.18^{+0.44}_{-0.33}$ & $5.58^{+0.67}_{-0.52}$ & $6.95^{+0.38}_{-0.28}$\\
             & Opt & $1.42^{+0.28}_{-0.23}$ & $5.43^{+0.32}_{-0.28}$ & $6.92^{+0.35}_{-0.25}$   \\
PKS\,2155-304 & HE & $-0.74^{+0.22}_{-0.19}$ & $4.39^{+0.30}_{-0.27}$ & $6.89^{+0.47}_{-0.47}$\\
             & Opt & $4.55^{+0.31}_{-0.24}$ &  $5.68^{+0.33}_{-0.27}$ &  $7.51^{+0.07}_{-0.13}$ \\
\hline
\end{tabular}
\caption{1$\sigma$ credible regions and the the maximum a posterior estimator for the hyper-parameters derived by the analysis adopting the AE$\times$CS kernel for the high-energy (HE) and the optical (Opt) light-curves. Flat uninformative or Jeffrey priors on the parameters were added to the likelihood function.}
\label{tab:AEC_hyp}
\end{table}

\begin{table}
\centering
\begin{tabular}{l|ccccc}
\hline
\hline
Source & band & $\ln A$   & $\ln \alpha$  & $\ln L$ & $\ln P$ \\
\hline 
PG\,1553+113 & HE & $-1.93^{+1.03}_{-0.52}$ & $-2.79^{+0.93}_{-1.34}$ & 
            $3.58^{+0.69}_{-0.94}$ & $6.67^{+0.04}_{-0.03}$ \\
             & Opt & $1.58^{+0.79}_{-0.38}$ & $-2.13^{+0.63}_{-1.05}$ & $3.85^{+0.37}_{-0.21}$ & $6.71^{+0.06}_{-0.04}$  \\
PKS\,2155-304 & HE & $-0.58^{+0.71}_{-0.31}$ & $-2.05^{+0.75}_{-1.20}$ &        
            $2.90^{+0.39}_{-0.70}$ & $6.44^{+0.13}_{-0.06}$ \\
             & Opt & $5.28^{+1.15}_{-0.64}$ & $-3.02^{+0.84}_{-1.27}$ & $4.20^{+0.57}_{-0.32}$ & $7.55^{+0.04}_{-0.07}$ \\
\hline
\end{tabular}
\caption{1$\sigma$ credible regions and the the maximum a posterior estimator for the hyper-parameters derived by the analysis adopting the RQ$\times$CS kernel for the high-energy (HE) and the optical (Opt) light-curves. Flat uninformative or Jeffrey priors on the parameters were added to the likelihood function.}
\label{tab:CRQ_hyp}
\end{table}

\begin{table}
\centering
\begin{tabular}{l|ccccc}
\hline
\hline
Source & band & $\ln A_{AE}$   & $\ln L$  & $\ln A_{CS}$ & $\ln P$ \\
\hline 
PG\,1553+113 & HE & $-2.76^{+0.52}_{-0.32}$ & $4.81^{+0.64}_{-0.48}$ & 
            $-3.20^{+1.31}_{-2.11}$ & $6.68^{+0.03}_{-0.03}$ \\
             & Opt & $1.17^{+0.40}_{-0.32}$ & $5.05^{+0.42}_{-0.34}$ & $-0.66^{+1.91}_{-12.0}$ & $6.69^{+0.05}_{-0.86}$  \\
PKS\,2155-304 & HE & $-0.97^{+0.25}_{-0.20}$ & $4.02^{+0.34}_{-0.29}$ &        
            $-11.6^{+5.9}_{-5.7}$ & $6.41^{+0.03}_{-0.04}$ \\
             & Opt & $4.70^{+0.47}_{-0.31}$ & $5.84^{+0.49}_{-0.33}$ & $-8.84^{+7.46}_{-7.54}$ & $6.09^{+1.04}_{-0.98}$ \\
\hline
\end{tabular}
\caption{1$\sigma$ credible regions and the the maximum a posterior estimator for the hyper-parameters derived by the analysis adopting the AE+CS kernel for the high-energy (HE) and the optical (Opt) light-curves. Flat uninformative or Jeffrey priors on the parameters were added to the likelihood function.}
\label{tab:CEXs_hyp}
\end{table}

\begin{table}
\centering
\begin{tabular}{l|cccccc}
\hline
\hline
Source & band & $\ln A_{RQ}$   & $\ln \alpha$ & $\ln L$ & $\ln A_{CS}$ & $\ln P$ \\
\hline 
PG\,1553+113 & HE & $-2.37^{+1.83}_{-0.62}$ & $-2.36^{+1.59}_{-2.41}$ & $3.36^{+0.75}_{-1.15}$ & $-2.89^{+1.39}_{-1.08}$ & $6.68^{+0.02}_{-0.02}$ \\
             & Opt & $1.27^{+1.26}_{-0.42}$ & $-1.48^{+0.81}_{-1.70}$ & $3.69^{+0.59}_{-0.21}$ & $0.26^{+1.44}_{-1.95}$ & $6.70^{+0.02}_{-0.03}$ \\
PKS\,2155-304 & HE & $-0.73^{+0.87}_{-0.33}$ & $-0.98^{+1.05}_{-1.70}$ & $3.19^{+0.30}_{-0.29}$ & $-4.52^{+3.00}_{-10.5}$ & $6.40^{+0.35}_{-0.84}$ \\
         & Opt & $6.11^{+2.08}_{-1.14}$ & $-4.04^{+1.29}_{-2.13}$ & $4.58^{+1.03}_{-0.56}$ & $-8.93^{+7.58}_{-7.53}$ & $6.15^{+1.03}_{-1.01}$ \\
\hline
\end{tabular}
\caption{1$\sigma$ credible regions and the maximum a posterior estimator for the hyper-parameters derived by the analysis adopting the RQ+CS kernel for the high-energy (HE) and the optical (Opt) light-curves. Flat uninformative or Jeffrey priors on the parameters were added to the likelihood function.}
\label{tab:CRQs_hyp}
\end{table}





\bibliography{scovino_GP}{}
\bibliographystyle{aasjournal}



\end{document}